\documentclass[reprint,aps,prb,superscriptaddress,showkeys]{revtex4-1}
\usepackage{graphicx}
\usepackage[pdftex]{hyperref}
\hypersetup{colorlinks=true,linkcolor=blue,citecolor=blue,urlcolor=blue}

\usepackage{color}
\usepackage{ulem}
\usepackage{version}
\usepackage{bm}
\usepackage{textcomp}
\usepackage{xfrac}

\DeclareMathAlphabet{\mathit}{OT1}{ptm}{m}{it}

\def \be {\begin{equation}}
\def \ee {\end{equation}}
\def \bea {\begin{eqnarray}}
\def \eea {\end{eqnarray}}

\def \bd  {\begin{details}}
\def \ed  {\end{details}}

\begin{document}

\excludeversion{details}


\title{Experimental evidence for Zeeman spin-orbit 
coupling \\ in layered antiferromagnetic conductors}  

\author{R. Ramazashvili}
\email[]{revaz@irsamc.ups-tlse.fr}
\affiliation{Laboratoire de Physique Th\'eorique, Universit\'e de Toulouse, CNRS, UPS, France}

\author{P.~D. Grigoriev}
\email[]{grigorev@itp.ac.ru}
\affiliation{L.~D. Landau Institute for Theoretical Physics, 142432 Chernogolovka, Russia}
\affiliation{National University of Science and Technology MISiS, 119049 Moscow, Russia}
\affiliation{P. N. Lebedev Physical Institute, 119991 Moscow, Russia}

\author{T. Helm}
\email[]{t.helm@hzdr.de}
\affiliation{Walther-Mei{\ss}ner-Institut, Bayerische Akademie der Wissenschaften, Walther-Mei{\ss}ner-Strasse 8, D-85748 Garching, Germany}
\affiliation{Physik-Department, Technische Universit{\"a}t M{\"u}nchen, D-$85748$ Garching, Germany}
\affiliation{Hochfeld-Magnetlabor Dresden (HLD-EMFL) and W\"{u}rzburg-Dresden Cluster of Excellence ct.qmat, Helmholtz-Zentrum Dresden-Rossendorf, 01328 Dresden, Germany}

\author{F. Kollmannsberger}
\affiliation{Walther-Mei{\ss}ner-Institut, Bayerische Akademie der Wissenschaften, Walther-Mei{\ss}ner-Strasse 8, D-85748 Garching, Germany}
\affiliation{Physik-Department, Technische Universit{\"a}t M{\"u}nchen, D-$85748$ Garching, Germany}

\author{M. Kunz}
\altaffiliation{Present address: 
TNG Technology Consulting GmbH, 85774 Unterf{\"o}hring, Germany}
\affiliation{Walther-Mei{\ss}ner-Institut, Bayerische Akademie der Wissenschaften, Walther-Mei{\ss}ner-Strasse 8, D-85748 Garching, Germany}
\affiliation{Physik-Department, Technische Universit{\"a}t M{\"u}nchen, D-$85748$ Garching, Germany}

\author{W. Biberacher}
\affiliation{Walther-Mei{\ss}ner-Institut, Bayerische Akademie der Wissenschaften, Walther-Mei{\ss}ner-Strasse 8, D-85748 Garching, Germany}

\author{E. Kampert}
\affiliation{Hochfeld-Magnetlabor Dresden (HLD-EMFL) and W\"{u}rzburg-Dresden Cluster of Excellence ct.qmat, Helmholtz-Zentrum Dresden-Rossendorf, 01328 Dresden, Germany}

\author{H. Fujiwara}
\affiliation{Department of Chemistry, Graduate School of Science, Osaka Prefecture University, Osaka 599-8531, Japan}

\author{A. Erb}
\affiliation{Walther-Mei{\ss}ner-Institut, Bayerische Akademie der Wissenschaften, Walther-Mei{\ss}ner-Strasse 8, D-85748 Garching, Germany}
\affiliation{Physik-Department, Technische Universit{\"a}t M{\"u}nchen, D-$85748$ Garching, Germany}

\author{J. Wosnitza}
\affiliation{Hochfeld-Magnetlabor Dresden (HLD-EMFL) and W\"{u}rzburg-Dresden Cluster of Excellence ct.qmat, Helmholtz-Zentrum Dresden-Rossendorf, 01328 Dresden, Germany}
\affiliation{Institut f\"{u}r Festk\"{o}rper- und Materialphysik, TU Dresden, 01062 Dresden, Germany}

\author{R. Gross}
\affiliation{Walther-Mei{\ss}ner-Institut, Bayerische Akademie der Wissenschaften, Walther-Mei{\ss}ner-Strasse 8, D-85748 Garching, Germany}
\affiliation{Physik-Department, Technische Universit{\"a}t M{\"u}nchen, D-$85748$ Garching, Germany}
\affiliation{Munich Center for Quantum Science and Technology (MCQST), D-80799 Munich, Germany}

\author{M.~V. Kartsovnik}
\email[]{mark.kartsovnik@wmi.badw.de}
\affiliation{Walther-Mei{\ss}ner-Institut, Bayerische Akademie der Wissenschaften, Walther-Mei{\ss}ner-Strasse 8, D-85748 Garching, Germany}

\date{\today}

\begin{abstract}

Most of solid-state spin physics arising from spin-orbit coupling, from fundamental phe\-no\-me\-na to industrial applications, relies on symmetry-protected degeneracies. So does the Zeeman spin-orbit coupling, expected to manifest itself in a wide range of antiferromagnetic conductors. Yet, experimental proof of this phenomenon has been lacking. Here, we demonstrate that the N\'eel state of the layered organic superconductor $\kappa$-(BETS)$_2$FeBr$_4$ shows no spin modulation of the Shubnikov-de Haas oscillations, contrary to its paramagnetic state. This is unambiguous evidence for the spin degeneracy of Landau levels, a direct manifestation of the Zeeman spin-orbit coupling. Likewise, we show that spin modulation is absent in electron-doped Nd$_{1.85}$Ce$_{0.15}$CuO$_4$, which evidences the presence of N\'eel order in this cuprate superconductor even at optimal doping. Obtained on two very different materials, our results demonstrate the generic character of the Zeeman spin-orbit coupling.
\end{abstract}

\maketitle

\section*{Introduction}
Spin-orbit coupling (SOC) in solids intertwines electron orbital motion with its spin, ge\-ne\-ra\-ting a variety of fundamental effects~\cite{Winkler,Dyakonov}. Commonly, SOC ori\-gi\-na\-tes from the Pauli term $\mathcal{H_P} = \frac{\hbar}{4m_e^2} {\bm \sigma} \cdot {\bf p} \times {\bm \nabla} V ({\bf r})$ in the electron Hamiltonian~\cite{LL-IV,Kittel}, where $\hbar$ is the Planck constant, $m_e$ the free electron mass, ${\bf p}$ the electron momentum, ${\bm \sigma}$ its spin and $V({\bf r})$ its potential energy depending on the position. Remarkably, N\'eel order may give rise to SOC of an entirely different nature, via the Zeeman effect~\cite{RevazPRL2008,RevazPRB2009-1}:
\be
\label{eq:Zeeman.SOC.1}
\mathcal{H}_\mathcal{Z}^{so} = - \frac{\mu_B}{2} \left[ g_\| ({\bf B}_\| \cdot {\bm \sigma}) + g_\perp ({\bf k}) ({\bf B}_\perp \cdot {\bm \sigma}) \right] ,
\ee
where $\mu_B$ is the Bohr magneton, ${\bf B}$ the magnetic field, while $g_\|$ and $g_\perp$ define the $g$-tensor components with respect to the N\'eel axis. In a purely transverse field ${\bf B}_\perp$, a hidden symmetry of a N\'eel antiferromagnet protects double degeneracy of Bloch eigenstates at a special set of momenta in the Brillouin zone(BZ)~\cite{RevazPRL2008,RevazPRB2009-1}: at such momenta, $g_\perp$ must vanish. The scale of $g_\perp$ is set by $g_\|$, which renders $g_\perp ({\bf k})$ substantially momentum dependent, and turns $\mathcal{H}^{so}_\mathcal{Z}$ into a veritable SOC~\cite{BraLuk1989,BraLuRa1989,RevazPRL2008,RevazPRB2009-1}. This coupling was predicted to produce unusual effects, such as spin degeneracy of Landau le\-vels in a purely transverse field ${\bf B}_\perp$~\cite{KabanovAlexandrov,RevazPRB2009-2} and spin-flip transitions, induced by an AC \textit{electric} rather than magnetic field~\cite{RevazPRB2009-2}.
Contrary to the textbook Pauli spin-orbit coupling, this mechanism does not require heavy elements. Being proportional to the applied magnetic field (and thus tunable!), it is bound only by the Néel temperature of the given material. In addition to its fundamental importance as a novel spin-orbit coupling mechanism, this phenomenon opens new possibilities for spin manipulation, much sought after in the current effort~\cite{Rashba-91,Shafiei-13, Berg-13} to harness electron spin for future spintronic applications. 
While the novel SOC mechanism may be relevant to a vast variety of antiferromagnetic (AF) conductors such as chromium, cup\-rates, iron pnic\-tides, he\-xa\-bo\-ri\-des, bo\-ro\-car\-bides, as well as organic and heavy-fermion compounds~\cite{RevazPRB2009-1}, it has not received an experimental confirmation yet.

Here, we present experimental evidence for the spin degeneracy of Landau levels in two very different layered conductors, using Shubnikov--de Haas (SdH) oscillations as a sensitive tool for quantifying the Zeeman effect~\cite{shoe84}. First, the organic superconductor $\kappa$-(BETS)$_2$FeBr$_4$ (hereafter $\kappa$-BETS)~\cite{fuji01} is employed for testing the theoretical predictions. The key features making this material a perfect model system for our purposes are (i) a simple quasi-two-dimensional (quasi-2D) Fermi surface and (ii) the possibility of tuning between the AF and paramagnetic (PM)  metallic states, both showing SdH oscillations, by a moderate magnetic field~\cite{fuji01,kono05}. We find that, contrary to what happens in the PM state, the angular dependence of the SdH oscillations in the AF state of this compound is \textit{not} modulated by the Zeeman splitting. We show that such a behavior is a na\-tu\-ral consequence  of commensurate N\'eel order giving rise to the Zeeman  SOC in the form of Eq.~(\ref{eq:Zeeman.SOC.1}). 

Having established the presence of the Zeeman SOC in an AF metal, we utilize this effect for probing the electronic state of Nd$_{2-x}$Ce$_x$CuO$_4$ (NCCO),
a prototypical example of electron-doped high-$T_c$ cuprate superconductors~\cite{armi10}. In these materials, superconductivity coexists with another symmetry-breaking phenomenon manifested in a Fermi-surface reconstruction as detected by angle-resolved photoemission spectroscopy (ARPES)~\cite{mats07,sant11,song17,he19} and SdH experiments~\cite{helm09,helm10,kart11,higg18}. The involvement of magnetism in this Fermi-surface reconstruction has been broadly debated~\cite{luke90,moto07,mang04a,saad15,yama03,kang05,daga05,yu07,dora18,das08,sach10,chak01,silv15,sach19,green20}. Here, we present detailed data on the SdH amplitude in optimally doped NCCO, tracing its variation over more than two orders of magnitude with changing the field orientation. The oscillation behavior is found to be very similar to that in $\kappa$-BETS. Given the crystal symmetry and the position of the relevant Fermi-surface pockets, this result is firm evidence for antiferromagnetism in NCCO. Our finding not only settles the controversy in electron-doped cuprate superconductors---but also clearly demonstrates the generality of the novel SOC mechanism.

\textit{Theoretical background} --- Before presenting the experimental results, we recapitulate the effect of Zeeman splitting on quantum oscillations. The superposition of the oscillations coming from the conduction subbands with opposite spins results in the well-known spin-reduction factor in the oscillation amplitude~\cite{shoe84}: $R_s = \cos \left(\pi \frac{g}{2} \frac{m}{m_{\mathrm{e}}} \right)$; here $m_{\mathrm{e}}$ and $m$ are, respectively, the free electron mass and the effective cyclotron mass of the relevant carriers. We restrict our consideration to the first-harmonic oscillations, which is fully sufficient for the description of our experimental results. In most three-dimensional (3D) metals, the dependence of $R_s$ on the field orientation is governed by the anisotropy of the cyclotron mass. At some field orientations $R_s$ may vanish, and the oscillation amplitude becomes zero. This {\it spin-zero} effect carries information about the renormalization of the product $gm$ relative to its free-electron value $2m_e$. For 3D systems, this effect is obviously not universal. For example, in the simplest case of a spherical Fermi surface, $R_s$ possesses no angular dependence whatsoever, hence no spin zeros. By contrast, in quasi-2D metals with their vanishingly weak interlayer dispersion such as in layered organic and cuprate conductors, spin zeros are~\cite{wosn92,kova93,meye95,uji02,kart16,RamshawNatPhys2011} a robust consequence of the monotonic increase of the cyclotron mass, $m \propto 1/\cos \theta$, with tilting the field by an angle $\theta$ away from the normal to the conducting layers. 
\begin{figure*}[tb]
	\centering
	\includegraphics[width=1\textwidth]{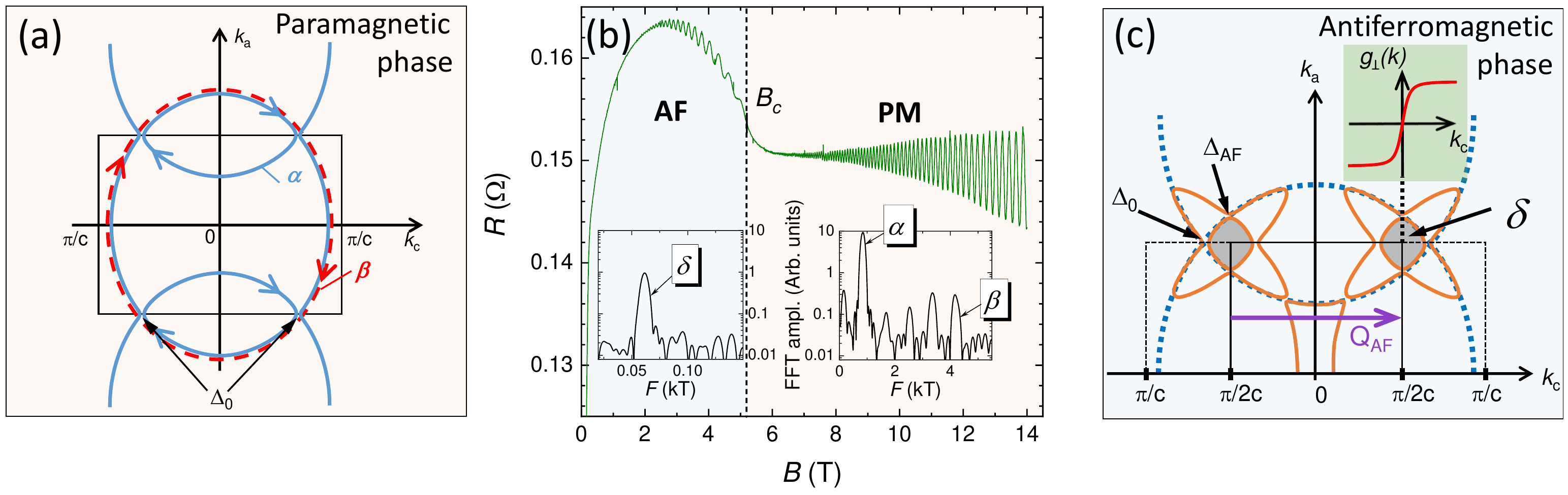}
	\caption{
		\textbf{2D Fermi surface of $\kappa$-BETS in the para- and antiferromagnetic phases. (a)} Fermi surface of $\kappa$-BETS in the PM state~\cite{uji01c,fuji01} (blue lines). The blue arrows show the classical cyclotron orbits $\alpha$ and the red arrows the large MB orbit $\beta$, which involves tunneling through four MB gaps $\Delta_{0}$ in a strong magnetic field. \textbf{(b)} Interlayer magnetoresistance of the $\kappa$-BETS sample, recorded at $T=0.5\,$K with field applied nearly perpendicularly to the layers ($\theta=2\,°\circ$). The vertical dash indicates the transition between the low-field AF and high-field PM states. The insets show the fast Fourier transforms (FFT) of the SdH oscillations for field windows $[2-5]\,$T and $[12-14]\,$T in the AF and PM state, respectively. \textbf{(c)} The BZ boundaries in the AF state with the wave vector $\mathbf{Q}_{\mathrm{AF}} = (\pi/c , 0)$ and in the PM state are shown by solid-black and dashed-black lines, respectively. The dotted-blue and solid-orange lines show, respectively, the original and reconstructed Fermi surfaces~\cite{kono05}. The shaded area in the corner of the magnetic BZ, separated from the rest of the Fermi surface by gaps $\Delta_0$ and $\Delta_{\text{AF}}$, is the $\delta$ pocket responsible for the SdH oscillations in the AF state.  The inset shows the function $g_\perp (\mathbf{k})$.}
	\label{BETS-FS}
\end{figure*}

In an AF metal, the $g$-factor may acquire a $\mathbf{k}$ dependence through the Zeeman SOC mechanism. It becomes particularly pronounced in the purely transverse geometry, i.e., for a magnetic field normal to the N\'eel axis. In this case, $R_s$ contains the factor $\bar{g}_{\perp}$ averaged over the cyclotron orbit, see Supplementary Information (hereafter SI) for details. As a result, the spin-reduction factor in a layered AF metal takes the form:
\be
\label{eq:R_s}
R_s = \cos \left[\frac{\pi}{\cos\theta} \frac{\bar{g}_\perp m_0}{2m_{\mathrm{e}}} \right]\,,
\ee
where $m_0 \equiv m(\theta = 0^{\circ})$.
Often, the Fermi surface is centered at a point ${\bf k^*}$, where the equality $g_\perp({\bf k^*}) = 0$ is protected by symmetry~\cite{RevazPRB2009-1}
 -- as it is for $\kappa$-BETS (see SI). 
 Such a ${\bf k^*}$ belongs to a line  node $g_\perp ({\bf k}) = 0$ crossing the Fermi surface. Hence, $g_\perp ({\bf k})$ changes sign along the Fermi surface, and $\bar{g}_\perp$ in Eq.~(\ref{eq:R_s}) vanishes by symmetry of $g_\perp({\bf k})$. Consequently, the quantum-oscillation amplitude is predicted to have \textit{no spin zeros}~\cite{KabanovAlexandrov}. For pockets with Fermi wave vector $k_F$ well below the inverse AF co\-he\-ren\-ce length $1/\xi$, $g_\perp({\bf k})$ can be described by the leading term of its expansion in ${\bf k}$. For such pockets, the present result was obtained in Refs.~\cite{KabanovAlexandrov,RevazPRB2009-2,RevazPRL2010}. According to our estimates in the SI, both in $\kappa$-BETS and in optimally doped NCCO ($x = 0.15$), the product $k_F \xi$ considerably exceeds unity. Yet, the quasi-classical consideration above shows that for $k_F \xi > 1$ the conclusion remains the same: $\bar{g}_\perp = 0$, see SI for the explicit theory. 
\begin{figure*}[tb]
	\centering
	\includegraphics[width=0.9\textwidth]{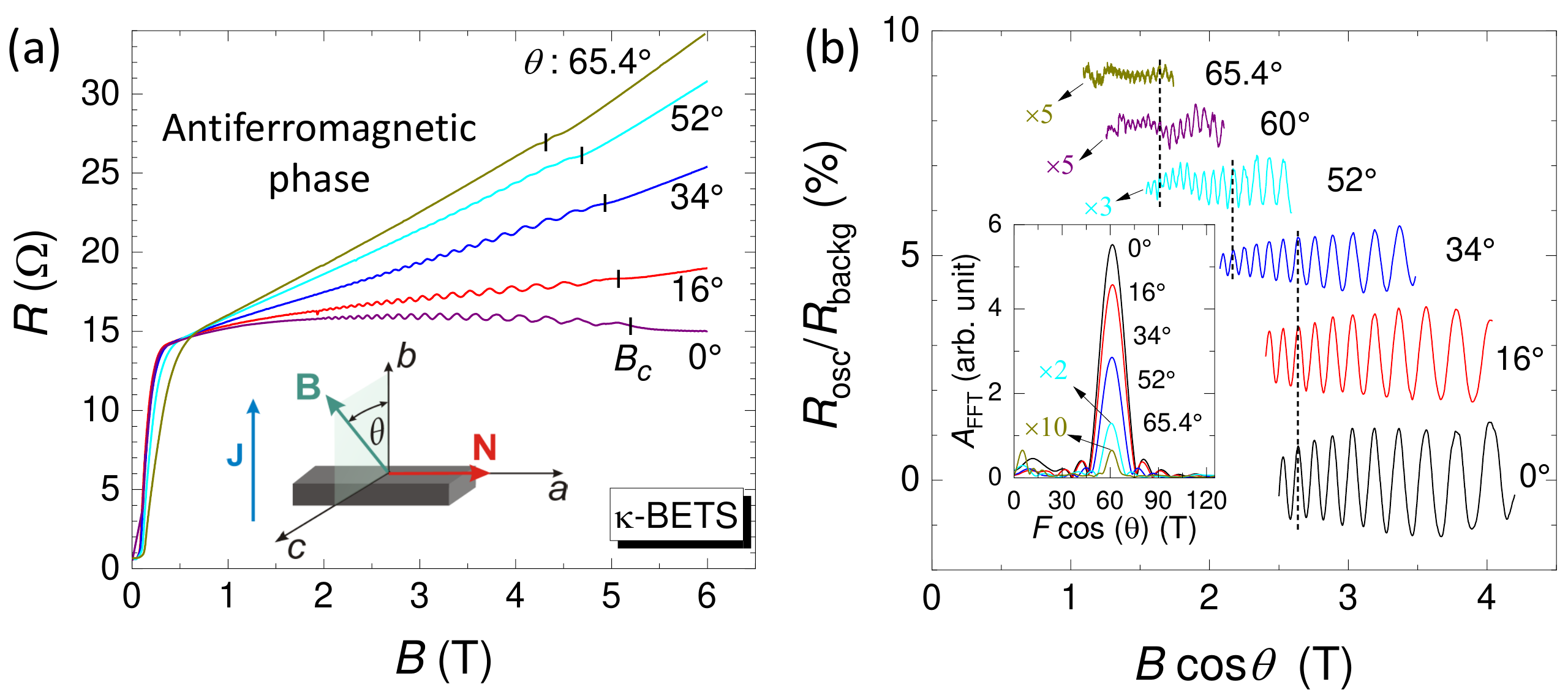}
	\caption{
		\textbf{SdH oscillations in the antiferromagnetic phase of $\kappa$-BETS}. \textbf{(a)} Examples of the field-dependent interlayer resistance at different field orientations, at $T = 0.42$\,K. The AF\,--\,PM transition field $B_c$ is marked by vertical dashes. \textbf{Inset:} the orientation of the current ${\bf J}$ and magnetic field ${\bf B}$ relative to the crystal axes and the N\'eel axis ${\bf N}$. \textbf{(b)} Oscillating component, normalized to the non-oscillating $B$-dependent resistance background, plotted as a function of the out-of-plane field component $B_{\perp}=B\cos\theta$. The curves corresponding to different tilt angles $\theta$ are vertically shifted for clarity. For $\theta \geq 52^{\circ}$ the ratio $R_{\text{osc}}/R_{\text{backg}}$ is multiplied by a constant factor, as indicated. The vertical dashed lines are drawn to emphasize the constant oscillation phase in these coordinates; \textbf{Inset:} FFT spectra of the SdH oscillations taken in the field window $[3-4.2]$\,T. The FFT amplitudes at $\theta = 52^{\circ}$ and $65.4^{\circ}$ are multiplied by a factor of 2 and 10, respectively.}
	\label{BETS-R(B)}
\end{figure*}

We emphasize that centering of the Fermi surface at a point ${\bf k^*}$ with $g_{\perp}({\bf k^*})=0$ -- such as a high-symmetry point of the magnetic BZ boundary~\cite{RevazPRB2009-1} -- is crucial for a vanishing of  $\bar{g}_\perp$. Otherwise, Zeeman SOC remains inert, as it does in AF CeIn$_3$, whose $d$ Fermi surface is centered at the $\Gamma$ point (see SI and Refs.~\cite{Sett01,Ebih04,Gork06}), and in quasi-2D EuMnBi$_2$, with its quartet of Dirac cones centered away from the magnetic BZ boundary~\cite{Mats18,Bori19}. With this, we turn to the experiment.

\section*{Results}

\textit{AF Organic Superconductor $\kappa$-(BETS)$_2$FeBr$_4$ ---}
This is a quasi-2D metal with conducting layers of BETS donor molecules, sandwiched between insulating FeBr$_4^-$-anion layers~\cite{fuji01}. The material has a centrosymmetric orthorhombic crystal structure (space group $Pnma$), with the $ac$ plane along the layers. The Fermi surface consists of a weakly warped cylinder and two open sheets, separated from the cylinder by a small gap $\Delta_0$ at the BZ boundary, as shown in Fig.~\ref{BETS-FS}~\cite{fuji01,kono05,uji01c}.

The magnetic properties of the compound are mainly go\-ver\-ned by five localized $3d$-electron spins per Fe$^{3+}$ ion in the insulating layers. Below $T_{\mathrm{N}} \approx 2.5$\,K, these  $S = 5/2$ spins are ordered antiferromagnetically, with the unit cell doubling along the $c$ axis and the staggered magnetization pointing along the $a$ axis~\cite{fuji01,mori02}. Above a critical magnetic field $B_c  \sim 2 - 5$\,T, dependent on the field orientation, antiferromagnetism gives way to a saturated PM state~\cite{kono04b}.

The SdH oscillations in the high-field PM state and in the N\'eel state are markedly different (see Fig.~\ref{BETS-FS}b). In the former, two dominant frequencies corresponding to a classical orbit $\alpha$ on the Fermi cylinder and to a large magnetic-breakdown (MB) orbit $\beta$ are found, in agreement with the predicted Fermi surface~\cite{uji01c,kono05}. The oscillation amplitude exhibits spin zeros as a function of the field strength and orientation, which is fairly well described by a field-dependent spin-reduction factor $R_s (\theta,B)$, with the $g$-factor $g = 2.0 \pm 0.2$ in the presence of an exchange field $B_J \approx -13$\,T, imposed by PM Fe$^{3+}$ ions on the conduction electrons~\cite{Cepas02,kart16}. In the SI, we provide further details of the SdH oscillation studies on $\kappa$-BETS.

Below $B_c$, in the AF state, new, slow oscillations at the frequency $F_{\delta} \approx 62$\,T emerge, indicating a Fermi-surface reconstruction~\cite{kono05}. The latter is associated with the folding of the original Fermi surface into the magnetic BZ, and $F_{\delta}$ is attributed to the new orbit $\delta$, see Fig.~\ref{BETS-FS}c. This orbit emerges due to the gap $\Delta_{\mathrm{AF}}$ at the Fermi-surface points, separated by the N\'eel wave vector $(\pi/c,0)$~\cite{Peierls.QTS}.

Figure~\ref{BETS-R(B)} shows examples of the field-dependent interlayer resistance of $\kappa$-BETS, recorded at $T=0.42$\,K, at different tilt angles $\theta$. The field was rotated in the plane normal to the N\'eel axis (crystallographic $a$ axis). In excellent agreement with previous reports~\cite{kono05,kono06}, slow oscillations with frequency $F_{\delta}=61.2\mathrm{\,T}/\cos\theta$ are observed below $B_c$, see inset in Fig.~\ref{BETS-R(B)}b. Thanks to the high crystal quality, even in this low-field region the oscillations can be traced over a wide angular range $|\theta| \leq 70^{\circ}$.

\begin{figure}[tb] 
	\centering
	\includegraphics[width=0.48\textwidth]{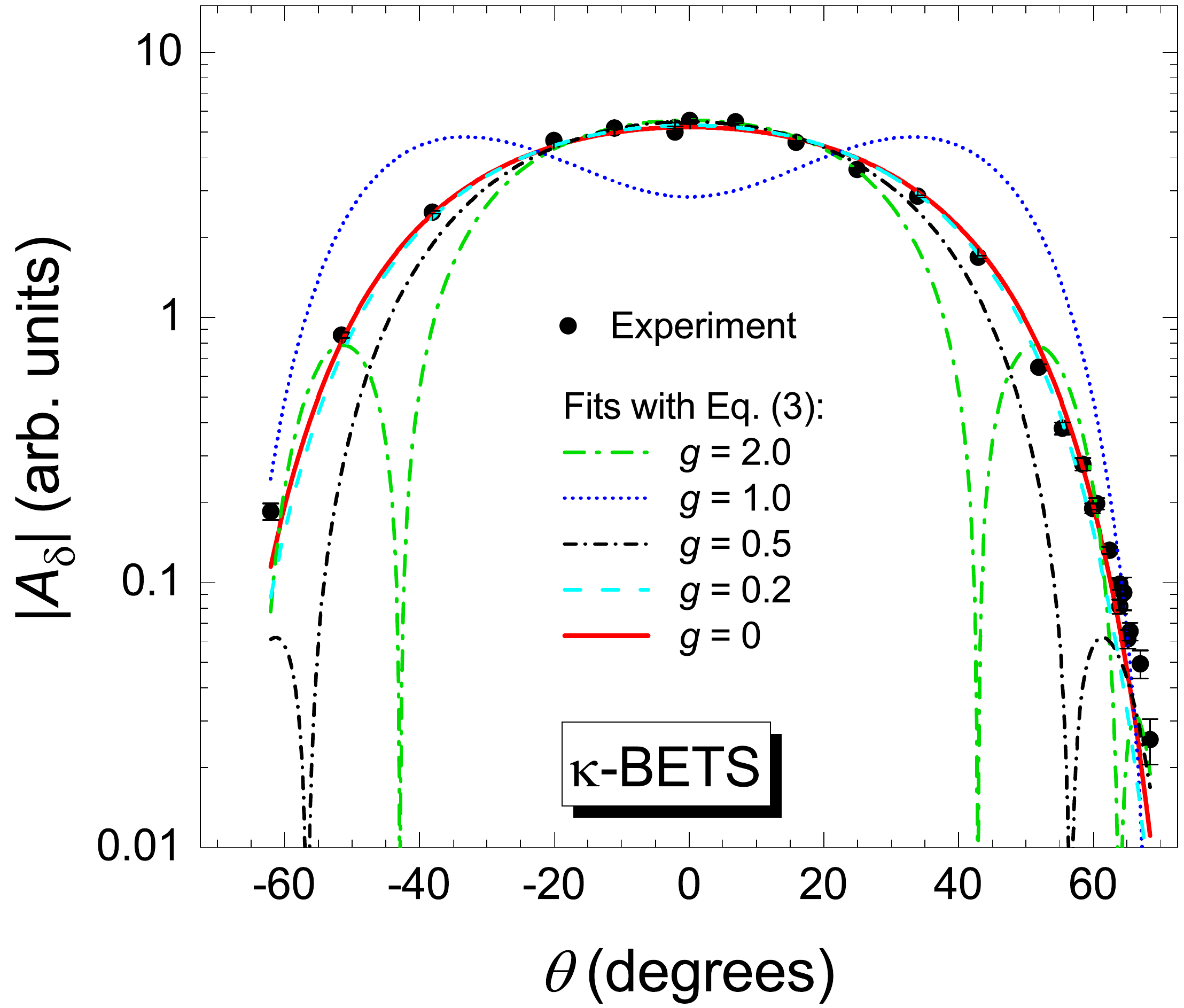}
	\caption{\textbf{Angular dependence of the SdH amplitude $A_\delta$ in the AF state of $\kappa$-BETS.} The lines are fits using Eq.~(\ref{LK}) with different values of the $g$-factor.}
	\label{BETS-A(theta)}
\end{figure}

The angular dependence of the $\delta$-oscillation amplitude $A_{\delta}$ is shown in Fig.~\ref{BETS-A(theta)}. The amplitude was determined by fast Fourier transform (FFT) of the zero-mean oscillating magnetoresistance component normalized to the monotonic $B$-dependent background, in the field window between $3.0$ and $4.2$\,T, so as to stay below $B_c (\theta)$ for all field orientations. The lines in Fig.~\ref{BETS-A(theta)} are fits using the Lifshitz-Kosevich formula for the SdH amplitude~\cite{shoe84}:
\be
A_{\delta} = A_0 \frac{m^2}{\sqrt{B}}R_{\mathrm{MB}}
\frac{\exp(-K m T_\mathrm{D}/B)}{\sinh(K m T/B)} R_s(\theta)\,,
\label{LK}
\ee
where $A_0$ is a field-independent prefactor, $B = 3.5$\,T (the midpoint of the FFT window in $1/B$ scale), $m$ the effective cyclotron mass ($m=1.1 m_{\mathrm{e}}$ at $\theta = 0^{\circ}$~\cite{kono05}, growing as $1/\cos\theta$ with tilting the field as in other quasi-2D metals~\cite{wosn96,kart04}), $K= 2\pi^2 k_B/\hbar e$, $T=0.42$\,K, $T_\mathrm{D}$ the Dingle tem\-pe\-ra\-tu\-re, and $R_{\mathrm{MB}}$ the MB factor. For $\kappa$-BETS, $R_{\mathrm{MB}}$ takes the form $R_{\mathrm{MB}} = \left[1-\exp\left(-\frac{B_0}{B\cos\theta}\right)\right] \left[1-\exp\left(-\frac{B_{\mathrm{AF}}}{B\cos\theta}\right)\right]$, with two cha\-rac\-te\-ris\-tic MB fields $B_0$ and $B_{\mathrm{AF}}$ associated with the gaps $\Delta_0$ and $\Delta_{\mathrm{AF}}$, respectively. The Zeeman splitting effect is encapsulated in the spin factor $R_s(\theta)$. In Eq.~(\ref{eq:Zeeman.SOC.1}), the geometry of our experiment implies $\mathbf{B}_\| = 0$, thus in the N\'eel state $R_s(\theta)$ takes the form
of Eq.~(\ref{eq:R_s}).

Excluding $R_s(\theta)$, the other factors in Eq.~(\ref{LK}) decrease monotonically with  increasing $\theta$. By contrast, $R_s(\theta)$ in Eq.~(\ref{eq:R_s}), generally, has an oscillating angular dependence. For $\bar{g}_{\perp} = g = 2.0$ found in the PM state~\cite{kart16}, Eq.~(\ref{eq:R_s}) yields two spin zeros, at $\theta \approx 43^{\circ}$ and $64^{\circ}$. Contrary to this, we observe \textit{no} spin zeros, but rather a monotonic decrease of $A_\delta$ by over two orders of magnitude as the field is tilted away from $\theta = 0^{\circ}$ to $\pm 70^{\circ}$, i.e., in the entire angular range where we observe the oscillations. The different curves in Fig.~\ref{BETS-A(theta)} are our fits using Eq.~(\ref{LK}) with $A_0$ and $T_\mathrm{D}$ as fit parameters, and dif\-fe\-rent va\-lu\-es of the $g$-factor. We used the MB field values $B_0 = 20$\,T and $B_{\mathrm{AF}} = 5$\,T. While the exact va\-lu\-es of $B_0$ and $B_{\mathrm{AF}}$ are unknown, they have virtually no effect on the fit quality, as we demonstrate in the SI. The best fit is achieved with $g = 0$, i.e. with an angle-independent spin factor $R_\mathrm{s} = 1$. The excellent agreement between the fit and the experimental data confirms the quasi-2D character of the electron conduction, with the $1/\cos\theta$ dependence of the cyclotron mass. 

Comparison of the curves in Fig.~\ref{BETS-A(theta)} with the data rules out $\bar{g}_\perp > 0.2$. Given the experimental error bars,  we cannot exclude a nonzero $\bar{g}_\perp \lesssim 0.2$, yet even such a small finite value would be in stark contrast with the textbook $g = 2.0$, found from the SdH oscillations in the high-field, PM state~~\cite{kart16}. Below we argue that, in fact, $\bar{g}_{\perp}$ in the N\'eel state is \textit{exactly} zero.

\begin{figure*}[tb]
	\centering
	\includegraphics[width=0.9\textwidth]{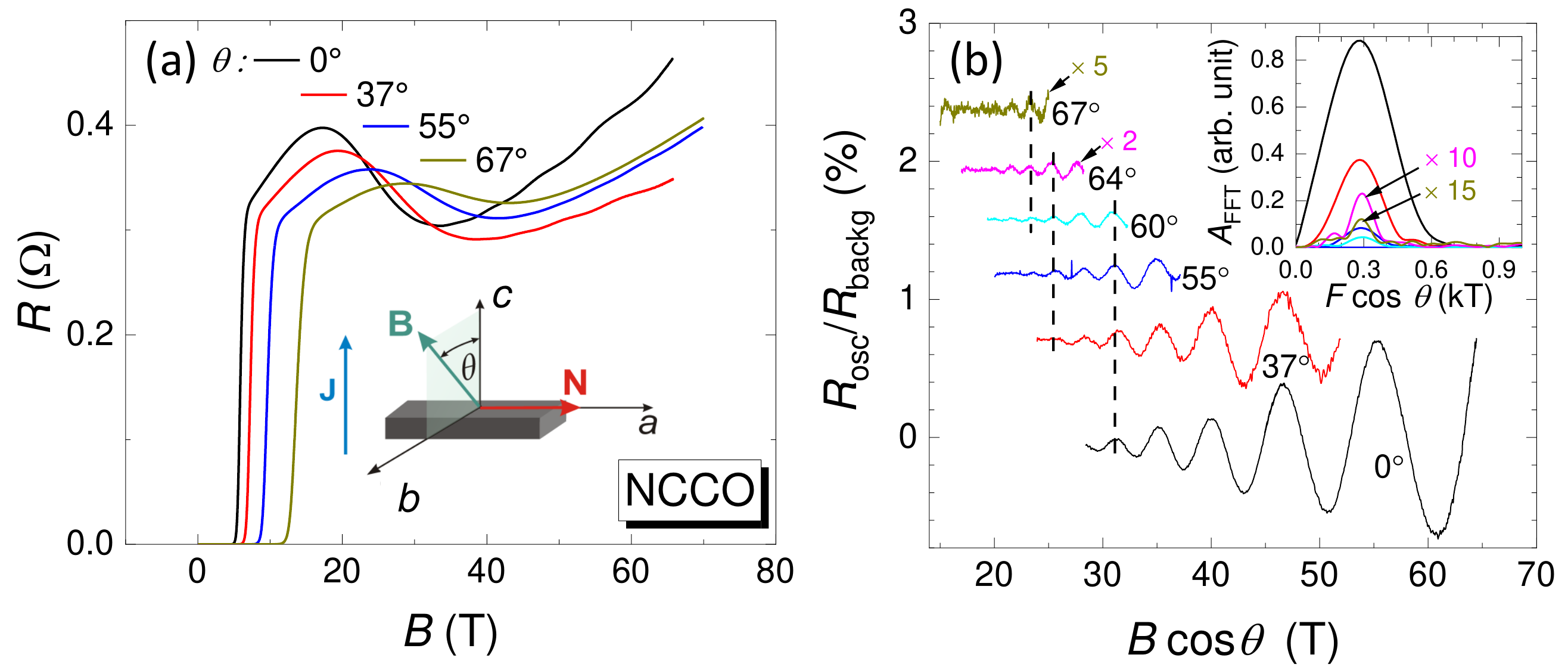}
	\caption{
		\textbf{Examples of MR and angle-dependent SdH oscillations in optimally doped NCCO. (a)} Examples of the $B$-dependent interlayer resistance at different field orientations, at $T=2.5$\,K; \textbf{(b)} oscillating component, normalized to the non-oscillating $B$-dependent resistance background, plotted as a function of the out-of-plane field component $B_{\perp}=B\cos \theta$. The curves corresponding to different tilt angles $\theta$ are vertically shifted for clarity. For $\theta = 64^{\circ}$ and $67^{\circ}$ the ratio $R_{\text{osc}}/R_{\text{backg}}$ is multiplied by a factor of 2 and 5, respectively. The vertical dashed lines are drawn to emphasize the constant oscillation phase in these coordinates; \textbf{Inset:} FFT spectra of the SdH oscillations taken in the field window 45 to 64\,T.}
	\label{NCCO-MR}
\end{figure*}
\textit{Optimally doped NCCO ---} This material has a body-centered tetragonal crystal structure (space group $I4/mmm$), where (001) conducting CuO$_2$ layers alternate with their insulating (Nd,Ce)O$_2$ counterparts~\cite{armi10}. Band-structure calculations~\cite{mass89, ande95} predict a holelike cylindrical Fermi surface, centered at the corner of the BZ. However, angle-resolved photoemission spectroscopy (ARPES)~\cite{armi02,mats07,sant11,song17} reveals a re\-con\-struc\-tion of this Fermi surface by a $(\pi / a , \pi / a)$ order. Moreover, magnetic quantum oscillations~\cite{helm10,kart11,higg18} show that the Fermi surface remains reconstructed even in the overdoped regime, up to the cri\-ti\-cal doping $x_c$ ($\approx 0.175$ for NCCO), where the superconductivity va\-ni\-shes~\cite{helm15}. The origin of this reconstruction remains unclear: while the $(\pi / a , \pi / a)$ periodicity is compatible with the N\'eel order observed in strongly underdoped NCCO, coexistence of antiferromagnetism and super\-con\-duc\-ti\-vi\-ty in electron-doped cuprates remains controversial. A number of neutron-scattering and muon-spin rotation studies~\cite{luke90,moto07,mang04a,saad15} have detected short-range N\'eel fluctuations, but no static order within the superconducting do\-ping range. However, other  neutron scattering~\cite{yama03,kang05} and magnetotransport~\cite{daga05, yu07,dora18} experiments have produced evi\-den\-ce of static or quasi-static AF order in superconducting samples at least up to optimal doping $x_{\mathrm{opt}}$. Alternative me\-cha\-nisms of the Fermi-surface reconstruction have been proposed, in\-clu\-ding a $d$-density wave~\cite{chak01}, a charge-density wave~\cite{silv15}, or coexistent topological and fluctuating short-range AF orders~\cite{sach19}.

To shed light on the relevance of antiferromagnetism to the electronic ground state of super\-con\-duc\-ting NCCO, we have studied the field-orientation dependence of the SdH oscillations of the interlayer resistance in an optimally doped, $x_{\mathrm{opt}} = 0.15$, NCCO crystal. The overall magnetoresistance behavior is illustrated in Fig.~\ref{NCCO-MR}(a). At low fields the sample is superconducting. Immediately above the $\theta$-dependent superconducting critical field the magnetoresistance displays a non-monotonic feature, which has already been reported for optimally doped NCCO in a magnetic field normal to the layers~\cite{helm09,li07a}. This anomaly correlates with an anomaly in the Hall resistance and has been associated with magnetic breakdown through the energy gap, created by the $(\pi/a , \pi / a)$-superlattice potential~\cite{helm15}. With increasing $\theta$, the anomaly shifts to higher fields, consistently with the expected increase of the breakdown gap with tilting the field.

\begin{figure}[tb]
	\centering
	\includegraphics[width=0.48\textwidth]{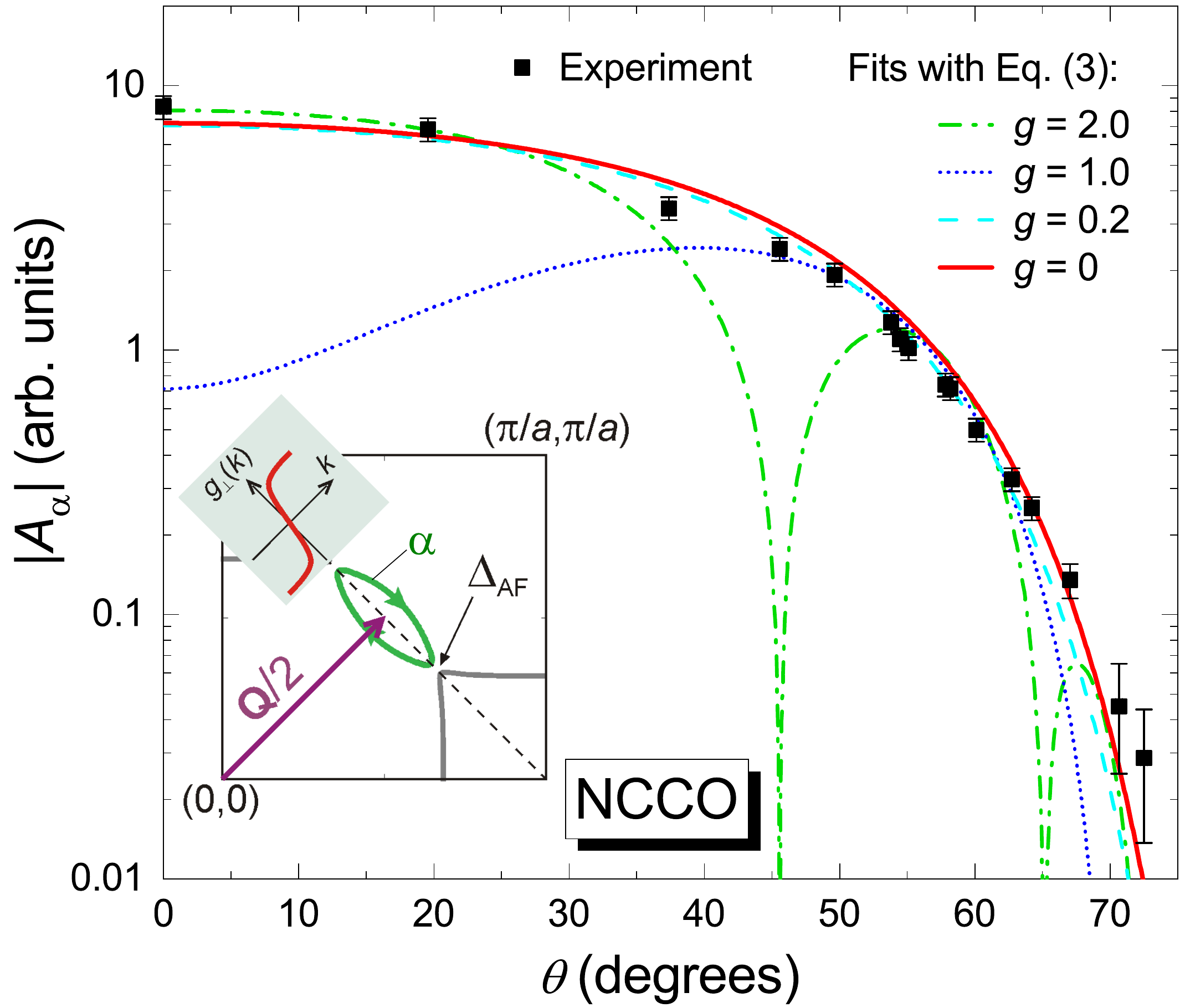}
	\caption{\textbf{Angular dependence of the SdH amplitude in  optimally doped NCCO.} The lines are fits using Eq.~(\ref{LK}) with different $g$-factor values. Inset: The first quadrant of the BZ with the Fermi-surface reconstructed by a superlattice potential with wave vector $\mathbf{Q}=(\pi/a,\pi/a)$. If this potential involves N\'eel order, the function $g_{\perp}(\mathbf{k})$ (red line in the inset) vanishes at the reduced BZ boundary (dashed line). The SdH oscillations are associated with the oval hole pocket $\alpha$ centered at $(\pi/2a,\pi/2a)$~\cite{helm09}.}
	\label{NCCO-A(theta)}
\end{figure}

SdH oscillations develop above about $30$\,T. Figure~\ref{NCCO-MR}(b) shows examples of the oscillatory component of the magnetoresistance, normalized to the field-dependent non-oscillatory background resistance $R_{\mathrm{backg}}$, determined by a low-order polynomial fit to the as-measured $R(B)$ dependence. In our conditions, $B\lesssim 65$\,T, $T = 2.5$\,K, the only  discernible contribution to the oscillations comes from the hole-like pocket $\alpha$ of the reconstructed Fermi surface~\cite{helm09}. This pocket is centered at the reduced BZ boundary, as shown in the inset of  Fig.~\ref{NCCO-A(theta)}. While magnetic breakdown creates large cyclotron orbits $\beta$ with the area equal to that of the unreconstructed Fermi surface, even in fields of 60-65\,T the fast $\beta$ oscillations are more than two orders of magnitude weaker than the $\alpha$ oscillations~\cite{kart11,helm15}.

The oscillatory signal is plotted in Fig.~\ref{NCCO-MR}(b) as a function of the out-of-plane field component $B_{\perp}=B\cos \theta$. In these coordinates the oscillation frequency remains constant, indicating that $F(\theta) = F(0^{\circ})/\cos \theta$ and thus confirming the quasi-2D character of the conduction. In the inset we show the respective FFTs plotted against the $\cos \theta$-scaled frequency. They exhibit a peak at $F\cos\theta = 294$\,T, in line with previous reports. The relatively large width of the FFT peaks is caused by the small number of oscillations in the field window $[45-64]$\,T. This restrictive choice is dictated by the requirement that the SdH oscillations be resolved over the whole field window at all tilt angles up to $\theta \approx 72^{\circ}$. In the SI we provide an additional analysis of the amplitude at fixed field values in the SI confirming the FFT results. Furthermore, in Fig.~\ref{NCCO-MR}(b) one can see that the phase of the oscillations is not inverted, and stays constant in the studied angular range. This is fully in line with the absence of spin zeros, see Eq.~(\ref{eq:R_s}).

The main panel of Fig.~\ref{NCCO-A(theta)} presents the angular dependence of the oscillation amplitude (symbols), in a field rotated in the $(ac)$ plane. The amplitude was determined by FFT of the data taken at $T=2.5$\,K in the field window $45$\,T $\leq B \leq 64$\,T. The lines in the figure are fits using Eq.~(\ref{LK}), for different $g$-factors. The fits were performed using the MB factor $R_{\mathrm{MB}} = [1-\exp(-B_0/B)]$~\cite{kart11,helm15}, the reported values for the MB field $B_0 =12.5$\,T, and the effective cyclotron mass $m(\theta=0^{\circ}) = 1.05 m_0$~\cite{helm15}, while taking into account  the $1/\cos\theta$ angular dependence of both $B_0$ and $m$. The prefactor $A_0$ and Dingle temperature $T_\mathrm{D}$ were used as fit parameters, yielding $T_\mathrm{D} = (12.6 \pm 1)$\,K, close to the value found  in the earlier experiment~\cite{helm15}. Note that, contrary to the hole-doped cuprate YBa$_2$Cu$_3$O$_{7-x}$, where the analysis of earlier ex\-pe\-ri\-ments~\cite{SebastianPRL2009,RamshawNatPhys2011} was complicated by the bilayer splitting of the Fermi cylinder~\cite{RamshawNatPhys2011}, the single-layer structure of NCCO poses no such difficulty.

Similar to $\kappa$-BETS, the oscillation amplitude in NCCO decreases by a factor of about $300$, with no sign of spin zeros as the field is tilted from $\theta =0^{\circ}$ to $72.5^{\circ}$. Again, this behavior is incompatible with the textbook value $g = 2$, which would have produced two spin zeros in the interval $0^{\circ} \leq \theta \leq 70^{\circ}$, see the green dash-dotted line in Fig.~\ref{NCCO-A(theta)}. A reduction of the $g$-factor to 1.0 would shift the first spin zero to about $72^{\circ}$, near the edge of our range (blue dotted line in Fig.~\ref{NCCO-A(theta)}). However, this would simultaneously suppress the amplitude at small $\theta$ by a factor of ten, contrary to our observations. All in all, our data rule out a constant $g > 0.2$.

\section*{Discussion}

In both materials, our data impose on the effective $g$-factor an upper bound of $0.2$. At first sight, one could simply view this as a suppression of the effective $g$ to a small nonzero value. However, below we argue that, in fact, our findings imply $\bar{g}_\perp = 0$ and point to the importance of the Zeeman SOC in both materials. The quasi-2D character of electron transport is crucial for this conclusion: as mentioned above, in three dimensions, the mere absence of spin zeros imposes no bounds on the $g$-factor. 

In $\kappa$-BETS the interplay between the crystal symmetry and the periodicity of the N\'eel state~\cite{RevazPRL2008, RevazPRB2009-1, RevazPRL2010} gua\-ran\-te\-es that $g_\perp ({\bf k})$ vanishes on the entire line $k_c = \pi/2c$ and is an odd function of $k_c - \pi/2c$, see the inset of Fig.~\ref{BETS-FS}c and Fig.~S2 in the SI. The $\delta$ orbit is centered on the line $k_c = \pi/2c$; hence $\bar{g}_\perp$ in Eq.~(\ref{eq:R_s}) vanishes, implying the absence of spin zeros, in agreement with our data. At the same time, quantum oscillations in the PM phase clearly reveal the Zeeman splitting of Landau levels with $g=2.0$. Therefore, we conclude that $\bar{g}_{\perp} =0$ is an intrinsic property of the N\'eel state.

In optimally doped NCCO, as already mentioned, the presence of a (quasi)static N\'eel order has been a subject of debate. However, if indeed present, such an order leads to $g_\perp ({\bf k}) = 0$ at the entire magnetic BZ boundary (see SI). For the hole pockets, producing the observed $F_{\alpha} \simeq 300$\,T oscillations, $\bar{g}_\perp = 0$ by symmetry of $g_\perp ({\bf k})$ (see inset of Fig.~\ref{NCCO-A(theta)} and Fig. S3 in SI). Such an interpretation requires that the relevant AF fluc\-tu\-a\-tions have frequencies below the cyclotron frequency in our experiment, $\nu_c \sim 10^{12}$\,Hz at 50\,T.

Finally, we address mechanisms -- other than Zeeman SOC of Eq.~(\ref{eq:Zeeman.SOC.1}) -- which may also lead to the absence of spin zeros. While such mechanisms do exist, we will show that none of them is relevant to the materials of our interest.

When looking for alternative explanations to our experimental findings, let us recall that,  generally, the effective $g$-factor may depend on the field orientation. This dependence may happen to compensate that of the quasi-2D cyclotron mass, $m/\cos\theta$, in the expression~(2) for the spin-reduction factor $R_s$, and render the latter nearly isotropic, with no spin zeros. Obviously, such a compensation requires a strong Ising anisotropy [$g(\theta=0^{\circ}) \gg g(\theta=90^{\circ})$] -- as found, for instance, in the heavy-fermion compound URu$_2$Si$_2$, with the values $g_c = 2.65 \pm 0.05$ and $g_{ab} = 0.0 \pm 0.1$ for the field along and normal to the $c$ axis, respectively~\cite{Alta12,Bast-19}. However, this scenario is irrelevant to both materials of our interest: In $\kappa$-BETS, a nearly isotropic $g$-factor, close to the free-electron value 2.0, was revealed by a study of spin zeros in the paramagnetic state~\cite{kart16}. In NCCO, the conduction electron $g$-factor may acquire anisotropy via an exchange coupling to Nd$^{3+}$ local moments. However, the low-temperature magnetic susceptibility of Nd$^{3+}$ in the basal plane is some 5 times larger than along the $c$ axis~\cite{Hund89,Dali93}. Therefore, the coupling to Nd$^{3+}$ may only \underline{increase} $g_{ab}$ relative to $g_c$, and thereby only \underline{enhance} the angular dependence of $R_s$ rather than cancel it  out. Thus, we are lead to rule out a $g$-factor anisotropy of crystal-field origin as a possible reason behind the absence of spin zeros in our experiments.

As follows from Eq.~(\ref{eq:R_s}), another possible reason for the absence of spin zeros is a strong reduction of the ratio $gm/2m_{\mathrm{e}}$. However, while \textit{some} renormalization of this ratio in me\-tals is commonplace, its dramatic suppression (let alone nullification) is, in fact, exceptional. Firstly, a vanishing mass would contradict $m/m_{\mathrm{e}} \approx 1$,  experimentally found in both materials at hand. On the other hand, a Landau Fermi-liquid renormalization~\cite{shoe84} $g \rightarrow g / (1 + G_0)$ would require a colossal Fermi-liquid parameter $G_0 \geq 10$, for which there is no evidence in NCCO, let alone $\kappa$-BETS with its already mentioned $g \approx 2$ in the paramagnetic state~\cite{kart16}.

A sufficient difference of the quantum-oscillation amplitudes and/or cyclotron masses for spin-up and spin-down Fermi surfaces might also lead to the absence of spin zeros. Some heavy-fermion compounds show strong spin polarization in magnetic field, concomitant with a substantial field-induced difference of the cyclotron masses of the two spin-split subbands~\cite{Harr-98,McCo-05}. 
As a result, for quantum oscillations in such materials, one spin amplitude considerably exceeds the other, and no spin zeros are expected. Note that this physics requires the presence of a very narrow conduction band, in addition to a broad one. In heavy-fermion compounds, such a band arises from the $f$ electrons, but is absent in both materials of our interest.

Another extreme example is given by the single fully polarized band in a ferromagnetic metal, where only one spin orientation is present, and spin zeros are obviously absent. Yet, no sign of ferro- or metamagnetism has been seen in either NCCO or $\kappa$-BETS. Moreover, in $\kappa$-BETS,the spin-zero effect has been observed in the paramagnetic state~\cite{kart16}, indicating that the quantum-oscillation amplitudes of the two spin-split subbands are comparable. However, for NCCO one may inquire whether spin polarization could render interlayer tunneling amplitudes for spin-up and spin-down different enough to lose spin zeros, especially in view of an extra contribution of Nd$^{3+}$ spins in the insulating layers to spin polarization. In the SI, we show that this is \textit{not} the case.

Thus, we are lead to conclude that the absence of spin zeros in the AF $\kappa$-BETS and in optimally doped NCCO is indeed a manifestation of the Zeeman SOC. Our explanation relies only on the symmetry of the N\'eel state and the location of the carrier pockets, while being insensitive to the me\-cha\-nism of the antiferromagnetism or to the orbital makeup of the relevant bands.

\section*{Methods}
Crystals of $\kappa$-(BETS)$_2$FeBr$_4$ were grown electrochemically and prepared for transport measurements as reported previously~\cite{kart16}. The interlayer ($I||b$) resistance was measured by the standard four-terminal a.c. technique using a low-frequency lock-in amplifier. Magnetoresistance measurements were performed in a superconducting magnet system at fields of up to $14\,$T. The samples were mounted on a holder allowing in-situ rotation of the sample around an axis perpendicular to the external field direction. The orientation of the crystal was defined by a polar angle $\theta$ between the field and the crystallographic $b$ axis (normal to the conducting layers).

Optimally doped single crystals of Nd$_{1.85}$Ce$_{0.15}$CuO$_4$, grown by the traveling solvent floating zone method, were prepared for transport measurements as reported previously~\cite{helm10}. Measurements of the interlayer ($I||c$) resistance were performed on a rotatable platform using a standard four-terminal a.c. technique at  frequencies of $30-70\,$kHz in a $70\,$T pulse-magnet system, with a pulse duration of 150 ms, at the Dresden High Magnetic Field Laboratory. The raw data were collected by a fast digitizing oscilloscope and processed afterwards by a digital lock-in procedure~\cite{helm09}. The orientation of the crystal was defined by a polar angle $\theta$ between the field and the crystallographic $c$ axis (normal to the conducting layers).

\section*{acknowledgments} 
It is our pleasure to thank G. Knebel, H. Sakai, I. Sheikin, and A. Yaresko for illuminating discussions. This work was supported by  HLD at HZDR, a member of the European Magnetic Field Laboratory (EMFL). P.~D.~G. acknowledges State Assignment \#0033-2019-0001 for financial support and the Laboratoire de Physique Th\'eorique, Toulouse, for the hos\-pi\-ta\-li\-ty du\-ring his visit. R.~G. acknowledges financial support from the German Research Foundation (Deutsche Forschungsgemeinschaft, DFG) via Germany's Excellence Strategy (EXC-2111-390814868). J.~W. acknowledges financial support from the DFG through the W\"{u}rzburg-Dresden Cluster of Excellence on Complexity and Topology in Quantum Matter - $ct.qmat$ (EXC 2147, project-id 390858490).   

\section*{Author Contributions}
T.H., F.K., M.K., E.K., W.B., and M.V.K performed the experiments and analyzed the data. R.R., P.D.G., and M.V.K. initiated the exploration and performed the theoretical analysis and interpretation of the experimental results. H.F. and A.E. provided high-quality single crystals. R.R., P.D.G., T.H., W.B., E.K., J.W., R.G., and M.V.K. contributed to the writing of the manuscript.

\section*{Data Availability}
The authors declare that all essential data supporting the findings of this study are available within the paper and its supplementary information. The complete data set in ASCII format is available from the corresponding authors upon reasonable request.

\section*{Additional Information}
\textbf{Supplementary Information:} Supplementary Information is available at [URL to be provided by the publisher].

\textbf{Competing interests:} The authors declare no competing interests.


\section*{References}
\begin{thebibliography}{9}
	\bibitem{Winkler}
	R. Winkler, Spin-orbit Coupling Effects in Two-dimensional Electron and Hole
	Systems, Springer Tracts in Modern Physics, Vol. 191, Berlin -- Heidelberg, 2003.
	\bibitem{Dyakonov}
	M. I. Dyakonov (Editor),
	Spin Physics in Semiconductors,
	Springer Series in Solid-State Sciences, Vol. 157, Berlin -- Heidelberg, 2017.
	\bibitem{LL-IV}
	V. B. Berestetski\u{\i}, E. M. Lifshitz and L. P. Pitaevski\u{\i},
	``Course of Theoretical Physics", Vol. 4,
	\textit{Quantum Electrodynamics}, Pergamon Press (1982).
	\bibitem{Kittel}
	C. Kittel, \textit{Quantum Theory of Solids}, John Wiley \& Sons, Inc., New York -- London (1963).
	\bibitem{RevazPRL2008} R. Ramazashvili,
	\textit{Kramers Degeneracy in a Magnetic Field and Zeeman
	 Spin-Orbit Coupling in Antiferromagnetic Conductors}, 
	Phys. Rev. Lett. \textbf{101}, 137202 (2008).
	\bibitem{RevazPRB2009-1} R. Ramazashvili,
	\textit{Kramers degeneracy in a magnetic field and Zeeman
	  spin-orbit coupling in antiferromagnetic conductors}, 
	Phys. Rev. B \textbf{79}, 184432 (2009).
	\bibitem{BraLuk1989}
	S. A. Brazovskii and I. A. Luk'yanchuk,
	\textit{Symmetry of electron states in antiferromagnets},
	Sov. Phys. JETP \textbf{69}, 1180-1184 (1989).
	\bibitem{BraLuRa1989}
	S. A. Brazovskii, I. A. Luk'yanchuk, and R. R. Ramazashvili,
	\textit{Electron paramagnetism in antiferromagnets},
	JETP Lett. \textbf{49}, 644-646 (1989).
	\bibitem{KabanovAlexandrov} V.V. Kabanov and A. S. Alexandrov, 
	\textit{Magnetic quantum oscillations in doped antiferromagnetic insulators}, 
	Phys. Rev. B \textbf{77}, 132403 (2008); \textbf{81}, 099907(E) (2010).
	\bibitem{RevazPRB2009-2} R. Ramazashvili,
	\textit{Electric excitation of spin resonance in antiferromagnetic conductors}, 
	Phys. Rev. B \textbf{80}, 054405 (2009).
	\bibitem{Rashba-91}
	E. I. Rashba and V. I. Sheka,
	\textit{"Electric-Dipole Spin Resonances", in Landau Level Spectroscopy},
	edited by G. Landwehr and E. I. Rashba (Elsevier, New York, (1991).
	\bibitem{Shafiei-13}
	M. Shafiei \textit{et al.}, 
	\textit{Resolving Spin-Orbit- and Hyperfine-Mediated Electric Dipole Spin Resonance in a Quantum Dot},
	Phys. Rev. Let. \textbf{110}, 107601 (2013).
	\bibitem{Berg-13}
	J. W. G. van den Berg \textit{et al.}, 
	\textit{Fast Spin-Orbit Qubit in an Indium Antimonide Nanowire},
	Phys. Rev. Let. \textbf{110}, 066806 (2013).
	\bibitem{shoe84} D. Shoenberg, 
	\textit{Magnetic Oscillations in Metals}, Cambridge University Press (1984).
	\bibitem{fuji01}
	H. Fujiwara \textit{et al.}, 
	\textit{A Novel Antiferromagnetic Organic Superconductor $\kappa$-(BETS)$_2$FeBr$_4$ [Where BETS = Bis(ethylenedithio)tetraselenafulvalene]}, 
	J. Am. Chem. Soc. {\bf 123}, 306-314 (2001).
	\bibitem{kono05}
	T. Konoike \textit{et al.}, 
	\textit{Fermi surface reconstruction in the magnetic-field-induced superconductor $\kappa$-(BETS)$_2$FeBr$_4$}, 
	Phys. Rev. B \textbf{72}, 094517 (2005).
	%
	\bibitem{armi10}
	N. P. Armitage, P. Fournier, and R. L. Greene, 
	\textit{Progress and perspectives on electron-doped cuprates}, 
	Rev. Mod. Phys. \textbf{82}, 2421-2487 (2010).
	\bibitem{sant11}
	A. F. Santander-Syro \textit{et al.}, 
	\textit{Two-Fermi-surface superconducting state and a nodal $d$-wave energy gap of the electron-doped Sm$_{1.85}$Ce$_{0.15}$CuO$_{4-\delta}$ cuprate superconductor},
	Phys. Rev. Lett. \textbf{106}, 197002 (2011).
	%
	\bibitem{song17}
	D. Song \textit{et al.}, 
	\textit{Electron Number-Based Phase Diagram of 
	Pr$_{1-x}$LaCe$_x$CuO$_{4-\delta}$ and Possible Absence of 
	Disparity between Electron- and Hole-Doped Cuprate Phase Diagrams}, 
	Phys. Rev. Lett. \textbf{118}, 137001 (2017).
	%
	\bibitem{he19}
	J.-F. He \textit{et al.}, 
	\textit{Fermi surface reconstruction in electron-doped cuprates without antiferromagnetic long-range order}, 
	Proc. Natl. Acad. Sci. USA {\bf 116}, 3449-3453 (2019).
	%
	\bibitem{mats07}
	H. Matsui \textit{et al.}, 
	\textit{Evolution of the pseudogap across the magnet-superconductor phase boundary of Nd$_{2-x}$Ce$_x$CuO$_4$}, 
	Phys. Rev. B \textbf{75}, 224514 (2007).
	%
	\bibitem{helm09}
	T. Helm \textit{et al.}, 
	\textit{Evolution of the Fermi surface of the electron-doped 
	high-temperature superconductor Nd$_{2-x}$Ce$_x$CuO$_4$ 
	revealed by Shubnikov - de Haas oscillations}, 
	Phys. Rev. Lett. \textbf{103}, 157002 (2009).
	%
	\bibitem{helm10}
	T. Helm \textit{et al.}, 
	\textit{Magnetic breakdown in the electron-doped cuprate superconductor 
	Nd$_{2-x}$Ce$_x$CuO$_4$: The reconstructed Fermi surface survives in the strongly overdoped regime}, 
	Phys. Rev. Lett. \textbf{105}, 247002 (2010).
	%
	\bibitem{kart11}
	M. V. Kartsovnik \textit{et al.}, 
	\textit{Fermi surface of the electron-doped cuprate superconductor 
	Nd$_{2-x}$Ce$_x$CuO$_4$ probed by high-field magnetotransport}, 
	New J. Phys. \textbf{13}, 015001 (2011).
	%
	\bibitem{higg18}
	J. S. Higgins \textit{et al.}, 
	\textit{Quantum oscillations from the reconstructed Fermi surface in electron-doped cuprate superconductors}, 
	New J. Phys. \textbf{20}, 043019 (2018).
	%
	\bibitem{das08}
	T. Das, R. S. Markiewicz, and A. Bansil, 
	\textit{Superconductivity and topological Fermi surface transitions in electron-doped cuprates near optimal doping}, 
	J. Phys. Chem. Solids \textbf{69}, 2963-2966 (2008).
	%
	\bibitem{sach10}
	S. Sachdev, 
	\textit{Where is the quantum critical point in the cuprate superconductors?}, 
	Phys. Status Solidi B \textbf{247}, 537-543 (2010).
	%
	\bibitem{chak01}
	S. Chakravarty, R. B. Laughlin, D. K. Morr, and C. Nayak,
	\textit{Hidden order in cuprates}, 
	Phys. Rev. B \textbf{63}, 094503 (2001).
	%
	\bibitem{silv15}
	E. H. da Silva Neto \textit{et al.}, 
	\textit{Charge ordering in the electron-doped 
	superconductor Nd$_{2–x}$Ce$_x$CuO$_4$}, 
	Science \textbf{347}, 282-285 (2015).
	%
	\bibitem{sach19}
	S. Sachdev,
	\textit{Topological order, emergent gauge fields, and Fermi surface reconstruction}, 
	Rep. Prog. Phys. \textbf{82}, 014001 (2019); S. Sachdev, H. D. Scammell, M. S. Scheurer, and G. Tarnopolsky,
	\textit{Gauge theory for the cuprates near optimal doping}, 
	Phys. Rev. B \textbf{99}, 054516 (2019).
	%
	\bibitem{luke90}
	G. M. Luke \textit{et al.}, 
	\textit{Magnetic order and electronic phase diagrams of electron-doped copper oxide materials}, 
	Phys. Rev. B \textbf{42}, 7981-7988 (1990).
	%
	\bibitem{moto07}
	E. M. Motoyama \textit{et al.}, 
	\textit{Spin correlations in the electron-doped high-transition-temperature superconductor Nd$_{2-x}$Ce$_x$CuO$_{4-\delta}$}, 
	Nature \textbf{445}, 186-189 (2007).
	%
	\bibitem{mang04a}
	P. K. Mang \textit{et al.}, 
	\textit{Phase decomposition and chemical inhomogeneity 
	in Nd$_{2-x}$Ce$_x$CuO$_{4\pm\delta}$}, 
	Phys. Rev. B \textbf{70}, 094507 (2004).
	%
	\bibitem{saad15}
	H. Saadaoui \textit{et al.}, 
	\textit{The phase diagram of electron-doped 
	La$_{2-x}$Ce$_x$CuO$_{4-\delta}$}, 
	Nat. Commun. \textbf{6}, 6041 (2015).
	%
	\bibitem{yama03}
	K. Yamada \textit{et al.}, 
	\textit{Commensurate Spin Dynamics in the Superconducting State of an Electron-Doped Cuprate Superconductor}, 
	Phys. Rev. Lett. \textbf{90}, 137004 (2003).
	%
	\bibitem{kang05}
	H. J. Kang \textit{et al.}, 
	\textit{Electronically competing phases and their magnetic field 
	dependence in electron-doped nonsuperconducting and superconducting Pr$_{0.88}$LaCe$_{0.12}$CuO$_{4\pm \delta}$}, 
	Phys. Rev. B \textbf{71}, 214512 (2005).
	%
	\bibitem{daga05}
	Y. Dagan \textit{et al.}, 
	\textit{Origin of the Anomalous Low Temperature Upturn in the Resistivity of the Electron-Doped Cuprate Superconductors}, 
	Phys. Rev. Lett. \textbf{94}, 057005 (2005).
	%
	\bibitem{yu07}
	W. Yu, J. S. Higgins, P. Bach, and R. L. Greene, 
	\textit{Transport evidence of a magnetic quantum phase transition in electron-doped high-temperature superconductors}, 
	Phys. Rev. B \textbf{76}, 020503(R) (2007).
	%
	\bibitem{dora18}
	A. Dorantes \textit{et al.}, 
	\textit{Magnetotransport evidence for irreversible spin reorientation in the collinear antiferromagnetic state of underdoped $Nd_{2-x}Ce_xCuO_4$}, 
	Phys. Rev. B \textbf{97}, 054430 (2018).
	%
	\bibitem{green20}
	R. L. Greene, P. R. Mandal, N. R. Poniatowski, and T. Sarkar, 
	\textit{The Strange Metal State of the Electron-Doped Cuprates},  
	Annu. Rev. Condens. Matter Phys. \textbf{11}, 213-229 (2020).
	%
	\bibitem{wosn92}
	J. Wosnitza \textit{et al.}, 
	\textit{de Haas–van Alphen studies of the organic superconductors $\alpha$-(ET)$_2$NH$_4$Hg(SCN)$_4$ and $\kappa$-(ET)$_2$Cu(NCS)$_2$ with ET = bis(ethelenedithio)-tetrathiafulvalene}, 
	Phys. Rev. B \textbf{45}, 3018-3025 (1992).
	%
	\bibitem{kova93}
	A. E. Kovalev, M. V. Kartsovnik, and N. D. Kushch, 
	\textit{Quantum and quasi-classical magnetoresistance 
	oscillations in a new organic metal (BEDT-TTF)$_2$TlHg(SeCN)$_4$}, 
	Solid State Commun. \textbf{87}, 705-708 (1993).
	%
	\bibitem{meye95}
	F. A. Meyer \textit{et al.},
	\textit{High-Field de Haas-Van Alphen Studies of 
	$\kappa$-(BEDT-TTF)$_2$Cu(NCS)$_2$}, 
	Europhys. Lett. \textbf{32}, 681-686 (1995).
	%
	\bibitem{uji02}
	S. Uji \textit{et al.}, 
	\textit{Fermi surface and internal magnetic field of the organic conductors 
	$\lambda$-(BETS)$_2$Fe$_x$Ga$_{1-x}$Cl$_4$}, 
	Phys. Rev. B \textbf{65}, 113101 (2002).
	%
	\bibitem{kart16}
	M.~V. Kartsovnik \textit{et al.}, 
	\textit{Interplay Between Conducting and Magnetic Systems in the Antiferromagnetic Organic Superconductor $\kappa$-(BETS)$_2$FeBr$_4$}, 
	J. Supercond. Nov. Magn. \textbf{29}, 3075-3080 (2016).
	\bibitem{RamshawNatPhys2011}
	B. J. Ramshaw \textit{et al.}, 
	\textit{Angle dependence of quantum oscillations in YBa$_2$Cu$_3$O$_{6:59}$ shows free-spin behaviour of quasiparticles}, 
	Nat. Phys. \textbf{7}, 234-238 (2011).
	%
	%
	\bibitem{RevazPRL2010}
	R. Ramazashvili, 
	\textit{Quantum Oscillations in Antiferromagnetic Conductors with Small Carrier Pockets},  
	Phys. Rev. Lett. \textbf{105}, 216404 (2010).
	%
	\bibitem{Sett01}
	R. Settai \textit{et al.}, 
	\textit{De Haas -- van Alphen studies of rare earth compounds}, 
	J. Magn. Magn. Mater. \textbf{140–144}, 1153-1154 (1995).
	%
	\bibitem{Ebih04}
	T. Ebihara, N. Harrison, M. Jaime, S. Uji, and J. C. Lashley,
	\textit{Emergent Fluctuation Hot Spots on the Fermi Surface of CeIn$_3$ in Strong Magnetic Fields}, 
	Phys. Rev. Lett. \textbf{93}, 246401 (2004).
	%
	\bibitem{Gork06}
	L. P. Gor’kov and P. D. Grigoriev,
	\textit{Antiferromagnetism and hot spots in CeIn$_3$}, 
	Phys. Rev. B \textbf{73}, 060401(R) (2006).
	%
	\bibitem{Mats18}
	H. Masuda \textit{et al.}, 
	\textit{Impact of antiferromagnetic order on Landau-level splitting of quasi-two-dimensional Dirac fermions in EuMnBi$_2$}, 
	Phys. Rev. B \textbf{98}, 161108(R) (2018).
	%
	\bibitem{Bori19}
	S. Borisenko \textit{et al.}, 
	\textit{Time-reversal symmetry breaking type-II Weyl state in YbMnBi$_2$}, 
	Nat. Commun. \textbf{10}, 3424 (2019).
	%
	\bibitem{uji01c}
	S. Uji \textit{et al.}, 
	\textit{Two-dimensional Fermi surface for the organic conductor $\kappa$-(BETS)$_2$FeBr$_4$}, 
	Physica B {\bf 298}, 557-561 (2001).
	%
	\bibitem{mori02}
	T. Mori and M. Katsuhara,
	\textit{Estimations of $\pi d$-interactions in organic conductors including magnetic anions},
	J. Phys. Soc. Jpn. \textbf{71}, 826-844 (2002).
	%
	\bibitem{kono04b}
	T. Konoike \textit{et al.}, 
	\textit{Magnetic-field-induced superconductivity in the antiferromagnetic organic superconductor $\kappa$-(BETS)$_2$FeBr$_4$}, 
	Phys. Rev. B \textbf{70}, 094514 (2004).
	%
	\bibitem{Cepas02}
	O.~C\'epas, R.~H.~McKenzie, and J.~Merino,
	\textit{Magnetic-field-induced superconductivity in layered organic molecular crystals with localized magnetic moments}, 
	Phys. Rev. B \textbf{65}, 100502(R) (2002).
	%
	\bibitem{Peierls.QTS}
	R. E. Peierls,
	\textit{Quantum Theory of Solids}, Oxford University Press (2001).
	%
	\bibitem{kono06}
	T. Konoike \textit{et al.}, 
	\textit{Anomalous {Magnetic}-{Field}-{Hysteresis} of {Quantum} {Oscillations} in $\kappa$-({BETS})$_2$FeBr$_4$}, 
	J. Low Temp. Phys. \textbf{142}, 531-534 (2006).
	%
	\bibitem{wosn96}
	J. Wosnitza,
	\textit{Fermi Surfaces of Low-Dimensional Organic Metals and Superconductors}, Springer Verlag, Berlin -- Heidelberg (1996).
	%
	\bibitem{kart04}
	M.~V. Kartsovnik, 
	\textit{High Magnetic Fields: A Tool for Studying Electronic Properties of Layered
Organic Metals}, 
	Chem. Rev. \textbf{104}, 5737-5782 (2004).
	%
	\bibitem{mass89}
	S. Massidda, N. Hamada, J. Yu, and A. J. Freeman,
	\textit{Electronic structure of Nd-Ce-Cu-O, a Fermi liquid superconductor},
	Physica C \textbf{157}, 571-574 (1989).
	%
	\bibitem{ande95}
	O. K. Andersen, A. I. Liechtenstein, O. Jepsen, and F. Paulsen,
	\textit{LDA energy bands, low-energy Hamiltonians, $t'$, $t''$, $t_\perp(k)$, and $J_\perp$},
	J. Phys. Chem. Solids \textbf{56}, 1573-1591 (1995).
	%
	\bibitem{armi02}
	N. P. Armitage \textit{et al.}, 
	\textit{Doping dependence of an n-type cuprate superconductor investigated by angle-resolved photoemission spectroscopy},
	Phys. Rev. Lett. \textbf{88}, 257001 (2002).
	%
	\bibitem{helm15}
	T. Helm \textit{et al.}, 
	\textit{Correlation between Fermi surface transformations and superconductivity in the electron-doped high-$T_c$ superconductor Nd$_{2-x}$Ce$_x$CuO$_4$}, 
	Phys. Rev. B \textbf{92}, 094501 (2015).
		%
\bibitem{li07a}
P. Li, F. F. Balakirev, and R. L. Greene, 
\textit{High-Field Hall Resistivity and Magnetoresistance of Electron-Doped Pr$_{2-x}$Ce$_x$CuO$_{4-\delta}$}, 
Phys. Rev. Lett. \textbf{99}, 047003 (2007). 

	%
	\bibitem{SebastianPRL2009}
	S. E. Sebastian \textit{et al.}, 
	\textit{Spin-Order Driven Fermi Surface Reconstruction Revealed by Quantum Oscillations in an Underdoped High $T_c$ Superconductor},
	Phys. Rev. Lett. \textbf{103}, 256405 (2009).
	%
	%
	\bibitem{Alta12}
	M. M. Altarawneh \textit{et al.}, 
	\textit{Superconducting Pairs with Extreme Uniaxial Anisotropy in URu$_2$Si$_2$},
	Phys. Rev. Lett. \textbf{108}, 066407 (2012).
	%
	\bibitem{Bast-19}
	G. Bastien \textit{et al.}, 
	\textit{Fermi-surface selective determination of the $g$-factor anisotropy in URu$_2$Si$_2$},
	Phys. Rev. B \textbf{99}, 165138 (2019).
	%
\bibitem{Hund89}
M. F. Hundley, J. D. Thompson, S-W. Cheong, Z. Fisk, and S. B. Oseroff,
\textit{Specific heat and anisotropic magnetic susceptibility of
Pr$_2$CuO$_4$, Nd$_2$CuO$_4$ and Sm$_2$CuO4 crystals},
Physica C \textbf{158}, 102-108 (1989). 
        %
\bibitem{Dali93}
Y. Dalichaouch, M. C. de Andrade, and M. B. Maple,
\textit{Synthesis, transport, and magnetic properties of
Ln$_{2 - x}$Ce$_x$Cu0$_{4 - y}$ single crystals (Ln = Nd, Pr, Sm)},
Physica C \textbf{218}, 309-315 (1993). 
	%
	\bibitem{Harr-98}
	N. Harrison, D. W. Hall, R. G. Goodrich, J. J. Vuillemin, and Z. Fisk, 
	\textit{Quantum Interference in the Spin-Polarized Heavy Fermion Compound
	CeB$_6$: Evidence for Topological Deformation of the Fermi Surface in Strong
	Magnetic Fields}, 
	Phys. Rev. Lett. {\bf 81}, 870 (1998).
	%
	\bibitem{McCo-05}
	A. McCollam \textit{et al.}, 
	\textit{Spin-dependent masses and field-induced quantum critical points}, 
	Physica B {\bf 359-361}, 1-8 (2005). 
\end{thebibliography}
\end{document}


\excludeversion{details}


\title{Experimental evidence for Zeeman spin-orbit 
coupling \\ in layered antiferromagnetic conductors}  

\author{R. Ramazashvili}
\email[]{revaz@irsamc.ups-tlse.fr}
\affiliation{Laboratoire de Physique Th\'eorique, Universit\'e de Toulouse, CNRS, UPS, France}

\author{P.~D. Grigoriev}
\email[]{grigorev@itp.ac.ru}
\affiliation{L.~D. Landau Institute for Theoretical Physics, 142432 Chernogolovka, Russia}
\affiliation{National University of Science and Technology MISiS, 119049 Moscow, Russia}
\affiliation{P.N. Lebedev Physical Institute, 119991 Moscow, Russia}

\author{T. Helm}
\altaffiliation{Present address:
Hochfeld-Magnetlabor Dresden (HLD-EMFL),
Helmholtz-Zentrum Dresden-Rossendorf, 01328 Dresden, Germany}
\affiliation{Walther-Mei{\ss}ner-Institut, Bayerische Akademie der Wissenschaften, Walther-Mei{\ss}ner-Strasse 8, D-85748 Garching, Germany}
\affiliation{Physik-Department, Technische Universit{\"a}t M{\"u}nchen, D-$85748$ Garching, Germany}

\author{F. Kollmannsberger}
\affiliation{Walther-Mei{\ss}ner-Institut, Bayerische Akademie der Wissenschaften, Walther-Mei{\ss}ner-Strasse 8, D-85748 Garching, Germany}
\affiliation{Physik-Department, Technische Universit{\"a}t M{\"u}nchen, D-$85748$ Garching, Germany}

\author{M. Kunz}
\altaffiliation{TNG Technology Consulting GmbH, 85774 Unterf{\"o}ring, Germany}
\affiliation{Walther-Mei{\ss}ner-Institut, Bayerische Akademie der Wissenschaften, Walther-Mei{\ss}ner-Strasse 8, D-85748 Garching, Germany}
\affiliation{Physik-Department, Technische Universit{\"a}t M{\"u}nchen, D-$85748$ Garching, Germany}

\author{W. Biberacher}
\affiliation{Walther-Mei{\ss}ner-Institut, Bayerische Akademie der Wissenschaften, Walther-Mei{\ss}ner-Strasse 8, D-85748 Garching, Germany}

\author{E. Kampert}
\affiliation{Hochfeld-Magnetlabor Dresden (HLD-EMFL) and W\"{u}rzburg-Dresden Cluster of Excellence ct.qmat, Helmholtz-Zentrum Dresden-Rossendorf, 01328 Dresden, Germany}

\author{H. Fujiwara}
\affiliation{Department of Chemistry, Graduate School of Science, Osaka Prefecture University, Osaka 599-8531, Japan}

\author{A. Erb}
\affiliation{Walther-Mei{\ss}ner-Institut, Bayerische Akademie der Wissenschaften, Walther-Mei{\ss}ner-Strasse 8, D-85748 Garching, Germany}
\affiliation{Physik-Department, Technische Universit{\"a}t M{\"u}nchen, D-$85748$ Garching, Germany}

\author{J. Wosnitza}
\affiliation{Hochfeld-Magnetlabor Dresden (HLD-EMFL) and W\"{u}rzburg-Dresden Cluster of Excellence ct.qmat, Helmholtz-Zentrum Dresden-Rossendorf, 01328 Dresden, Germany}
\affiliation{Institut f\"{u}r Festk\"{o}rper- und Materialphysik, TU Dresden, 01062 Dresden, Germany}

\author{R. Gross}
\affiliation{Walther-Mei{\ss}ner-Institut, Bayerische Akademie der Wissenschaften, Walther-Mei{\ss}ner-Strasse 8, D-85748 Garching, Germany}
\affiliation{Physik-Department, Technische Universit{\"a}t M{\"u}nchen, D-$85748$ Garching, Germany}
\affiliation{Munich Center for Quantum Science and Technology (MCQST), D-80799 Munich, Germany}

\author{M.~V. Kartsovnik}
\email[]{mark.kartsovnik@wmi.badw.de}
\affiliation{Walther-Mei{\ss}ner-Institut, Bayerische Akademie der Wissenschaften, Walther-Mei{\ss}ner-Strasse 8, D-85748 Garching, Germany}

\date{\today}

\section*{Supplementary Information}

\setcounter{figure}{0}
\makeatletter 
\renewcommand{\thefigure}{S\@arabic\c@figure}
\makeatother
\section{Spin-reduction factor $R_s(\theta)$ in a magnetic field perpendicular to the N\'{e}el axis} \label{sec:dF}

For a quasi-2D metal, the phase $\varphi$ of the first quantum-oscillation harmonic (i.e., of the fundamental frequency oscillations) is, up to a constant,
$
\varphi = \frac{F}{B} = \frac{\hbar}{e} \frac{\mathcal{F}}{B \cos \theta}\,,
$
where $\theta$ is the tilt angle between the field and the normal to the conducting plane, $F$ stands for the oscillation frequency at $\theta = 0^{\circ}$, and $\mathcal{F}$ the corresponding Fermi-surface area. The Zeeman effect splits the spin-degenerate Fermi surface, breaking up $\mathcal{F}$ into $\mathcal{F}_+$ and $\mathcal{F}_-$, with $\delta \mathcal{F} = \mathcal{F}_+ - \mathcal{F}_- \propto B$. Adding the two harmonic oscillations at close frequencies $F_+$ and $F_-$ is equivalent to a single oscillation at frequency $F$, with an amplitude modulated by the spin-reduction factor
\be
\label{eq:Rs.via.F}
R_s ( \theta ) = \cos  \left[ \frac{\hbar}{2e} \frac{\delta \mathcal{F}}{B \cos \theta} \right] .
\ee
\begin{figure}[b] 
	\center
	\includegraphics[width = .35 \textwidth]{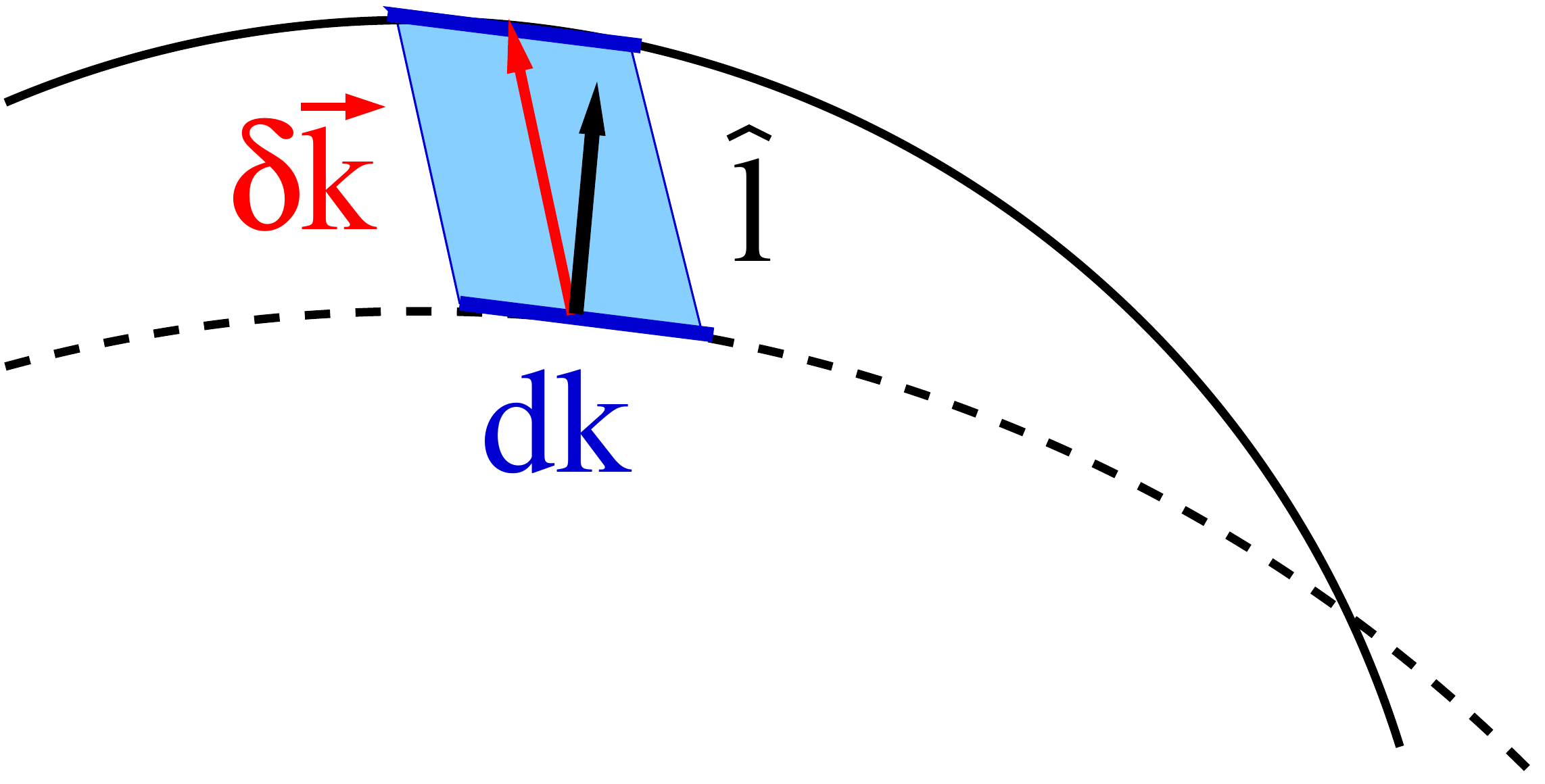}
	\caption{\textbf{Fermi surface of a two-dimensional conductor in zero field (dashed line) and in a transverse field ${\bf B}_\perp$ (solid line).} Upon turning on ${\bf B}_\perp$, a Fermi-surface element $dk$ shifts by a small momentum $\delta {\bf k}$ (see the main text), adding the shaded trapezoid of the area $d \mathcal{F} = dk (\delta {\bf k} \cdot \hat{\bf l}_{\bf k})$ to the area enclosed by the Fermi surface, where $\hat{\bf l}_{\bf k} = {\bm \nabla}_{\bf k} \mathcal{E}({\bf k}) / | {\bm \nabla}_{\bf k} \mathcal{E}({\bf k}) |$ is the local unit vector, normal to the Fermi surface. The total variation of the Fermi-surface area is given by integrating $d \mathcal{F}$ over the Fermi surface, as explained in the main text.}
	\label{fig:dF}
\end{figure}

For the field perpendicular to the N\'eel axis, ${\bf B}={\bf B}_\perp$, Eq.~(1) of the main text yields the single-particle Hamiltonian: $\mathcal{H} = \mathcal{E}({\bf k}) -  \frac{1}{2} \mu_B g_\perp({\bf k}) ({\bf B}_\perp \cdot {\bm \sigma})$, where $\mathcal{E}({\bf k})$ is the zero-field dispersion near the Fermi surface, and ${\bm \sigma} = (\sigma^x , \sigma^y , \sigma^z)$ is a vector composed of the three Pauli matrices. Upon turning on the field, a given point ${\bf k}$ of the Fermi surface undergoes a small shift $\delta {\bf k}$ such that $\delta {\bf k} \cdot {\bm \nabla}_{\bf k} \mathcal{E}({\bf k}) = \pm \frac{1}{2} \mu_B g_\perp({\bf k}) B_\perp$, where the $\pm$ signs correspond to the `up' and `down' spin projections on ${\bf B}_\perp$ and to the subscript of the resulting Fermi-surface areas $\mathcal{F}_\pm$. As shown in Fig.~\ref{fig:dF}, upon the shift by $\delta {\bf k}$ an element $dk$ of the Fermi surface contributes the shaded area $dk (\delta {\bf k} \cdot \hat{\bf l}_{\bf k}) = \pm \frac{1}{2} \mu_B B_\perp dk g_\perp({\bf k}) / | {\bm \nabla}_{\bf k} \mathcal{E}({\bf k}) |$ to the variation of the total area, enclosed by the Fermi surface. Here, $\hat{\bf l}_{\bf k} = {\bm \nabla}_{\bf k} \mathcal{E}({\bf k}) / | {\bm \nabla}_{\bf k} \mathcal{E}({\bf k}) |$ is the local unit vector, normal to the Fermi surface. Therefore, to linear order in $B_\perp$, the areas $\mathcal{F}_\pm$ of the two spin-split Fermi surfaces differ by
\be
\label{eq:dF}
\delta \mathcal{F} =
\mathcal{F}_+ -  \mathcal{F}_- = \mu_B B_\perp \oint_{FS} \frac{dk g_\perp({\bf k})}{|\nabla_{\bf k} \mathcal{E}({\bf k}) |} ,
\ee
where the line integral is taken along the zero-field Fermi surface. It is convenient to introduce the transverse $g$-factor, averaged over the Fermi surface:
\be
\label{eq:g.average}
\bar{g}_\perp = \oint_{FS} \frac{\hbar^2 dk}{2 \pi m} \frac{g_\perp({\bf k})}{|\nabla_{\bf k} \mathcal{E}({\bf k}) |} ,
\ee
where
\be
\label{eq:m.average}
m = \frac{1}{2 \pi} \oint_{FS} \frac{\hbar^2 dk}{|\nabla_{\bf k} \mathcal{E}({\bf k}) |}
\ee
is the ($\theta = 0^{\circ}$) cyclotron mass. Substituting $\mu_B  = \frac{e\hbar}{2m_{\mathrm{e}}}$ into Eq.~(\ref{eq:dF}), we combine it with Eq.~(\ref{eq:Rs.via.F})  to arrive at the expression for the spin-reduction factor:
\be
\label{eq:R_s2}
R_s (\theta) = \cos \left[ \frac{\pi}{\cos \theta} \frac{\bar{g}_\perp m}{2m_{\mathrm{e}}} \right] ,
\ee
that is Eq.\,(2) of the main text. For a momentum-independent $g_\perp$, Eq.~(\ref{eq:R_s2}) matches the textbook expression~\cite{shoe84} for the spin-reduction factor in two dimensions. 

\section{Symmetry analysis of the N\'eel states of
$\kappa$-(BETS) and NCCO}

\subsection{Symmetry analysis of the N\'eel state of
$\kappa$-(BETS)$_2\text{FeBr}_4$ in a transverse field}
\label{Sec:BETS-symmetry}

The existence of a special set of momenta in the Brillouin zone (BZ), where Bloch eigenstates of a N\'eel antiferromagnet remain degenerate in transverse magnetic field, is a general phenomenon. However, the precise geo\-met\-ry of this set depends on the interplay between the periodicity of the N\'eel order and the symmetry of the underlying crystal lattice~\cite{RevazPRL2008,RevazPRB2009-1}. Here, we describe this set for $\kappa$-(BETS)$_2$FeBr$_4$, hereafter referred to as $\kappa$-BETS.

Upon transition from the paramagnetic to N\'eel state, the lattice period of $\kappa$-BETS along the $c$ axis doubles, and the symmetry of the paramagnetic state with respect to both the time reversal $\hat{\theta}$ and the elementary translation $\hat{{\bf T}}_c$ along the $c$ axis is broken. Yet, the product $\hat{\theta}\hat{\bf T}_c$ remains a symmetry operation, along with the spin rotation $\hat{\bf U}_{\bf n} (\phi)$ around the N\'eel axis ${\bf n}$ by an ar\-bit\-ra\-ry angle $\phi$.

\begin{figure}[tb] 
\center
\includegraphics[width = .48 \textwidth]{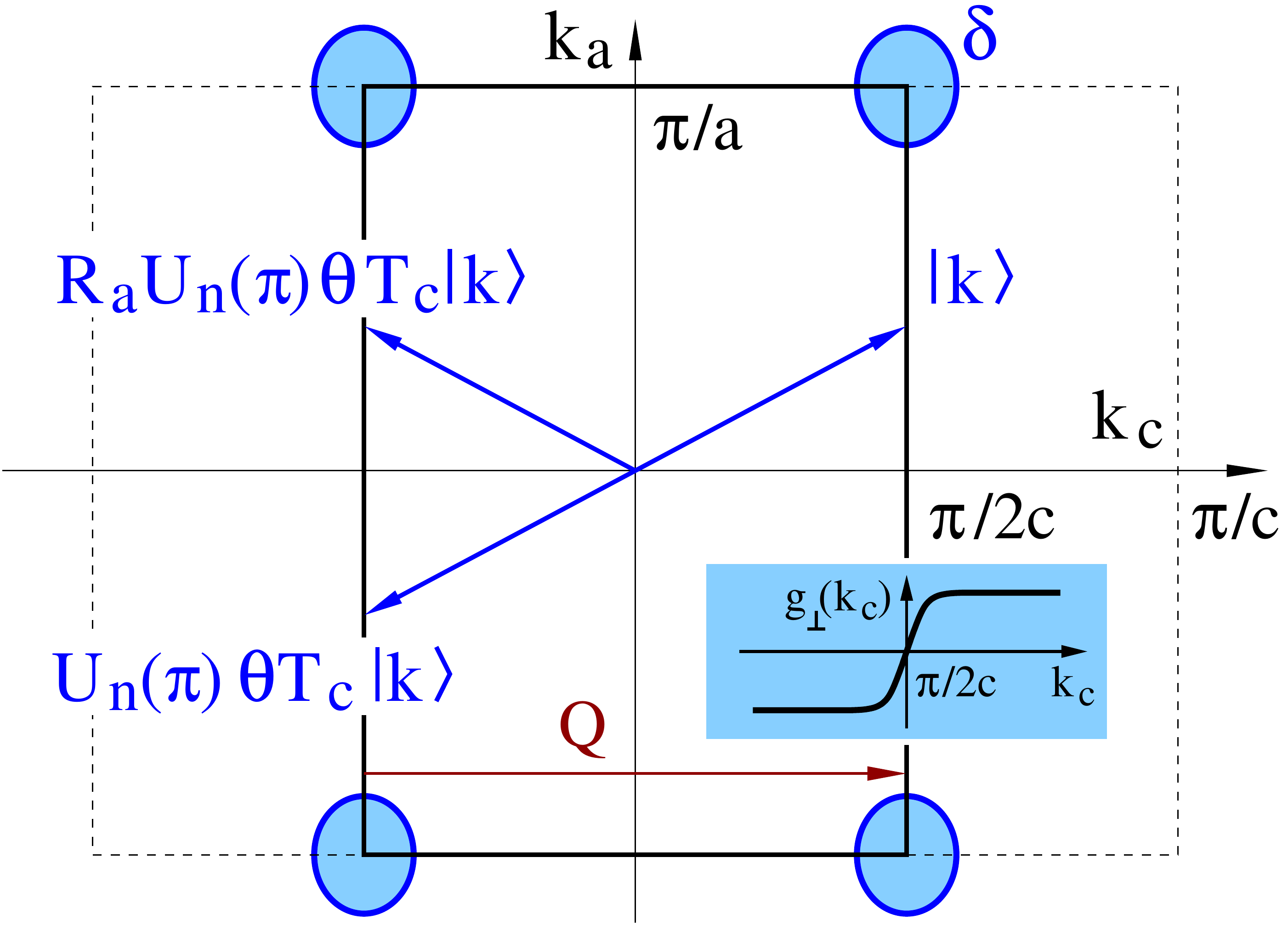}
\caption{\textbf{Schematic view of the BZ of $\kappa$-BETS.} The BZ in its paramagnetic state (dashed line) and in the N\'eel state with wave vector ${\bf Q} = (\pi/c , 0)$ (solid line). The $\delta$ pocket is centered at the corner ${\bf k} = (\pm \pi/2c , \pm \pi / a) $ of the magnetic BZ. The arrows show an exact Bloch eigenstate $| {\bf k} \rangle$ at a wave vector ${\bf k}$ on the vertical segment $k_c = \pi/2c$ of the magnetic BZ boundary -- and its symmetry partners ${\bf U}_{\bf n} (\pi) {\theta} {\bf T}_c | {\bf k} \rangle$ and
${\bf R}_a {\bf U}_{\bf n} (\pi) {\theta} {\bf T}_c  | {\bf k} \rangle$. The orthogonality $\langle {\bf k} | {\bf R}_a {\bf U}_{\bf n} (\pi) {\theta} {\bf T}_c | {\bf k} \rangle = 0$ implies that $g_\perp ({\bf k})$ vanishes on the segment $k_c = \pi/2c$. The inset illustrates the $g_\perp ({\bf k})$ being an odd function of $k_c - \pi/2c$.}
\label{fig:BETS-sym}
\end{figure}

Applied transversely to ${\bf n}$, a magnetic field breaks the symmetry with respect to both $\hat{\theta}\hat{\bf T}_c$ and $\hat{\bf U}_{\bf n} (\phi)$; however, $\hat{\bf U}_{\bf n} (\pi) \hat{\theta} \hat{\bf T}_c$ remains a symmetry operation~\cite{RevazPRL2008,RevazPRB2009-1}. It maps a Bloch eigenstate $| {\bf k} \rangle$ at wave vector ${\bf k}$ onto a degenerate orthogonal eigenstate $ \hat{\bf U}_{\bf n} (\pi) \hat{\theta} \hat{\bf T}_c | {\bf k} \rangle$ at wave vector $-{\bf k}$, as shown in Fig.~\ref{fig:BETS-sym}. Upon combination with reflection $\hat{\bf R}_a$: $(k_c , k_a) \rightarrow (k_c , - k_a)$, the resulting symmetry operation $\hat{\bf R}_a \hat{\bf U}_{\bf n} (\pi) \hat{\theta} \hat{\bf T}_c$ maps $| {\bf k} \rangle$ at wave vector ${\bf k} = (k_c , k_a)$ onto a degenerate orthogonal eigenstate $\hat{\bf R}_a \hat{\bf U}_{\bf n} (\pi) \hat{\theta} \hat{\bf T}_c  | {\bf k} \rangle$ at wave vector $(-k_c , k_a)$~\cite{RevazPRL2008,RevazPRB2009-1}.

For an arbitrary ${\bf k} = (k_c , k_a)$ at the vertical segment $k_c = \pi/2c$ of the magnetic BZ boundary, the wave vectors $(-k_c , k_a)$ and $(k_c , k_a)$ differ by the reciprocal wave vector ${\bf Q} = (\pi/c , 0)$ of the N\'eel state; in the nomenclature of the magnetic BZ, they are one and the same vector. The degeneracy of such a $| {\bf k} \rangle$ with $\hat{\theta} \hat{\bf T}_c \hat{\bf R}_a \hat{\bf U}_{\bf n} (\pi) | {\bf k} \rangle$ means that $g_\perp ({\bf k})$ vanishes at the entire segment $k_c = \pi/2c$.

The $\delta$ pocket is centered at $(\pm \pi/2c , \pm \pi/a)$ and is symmetric with respect to reflection around the line $k_c = \pm \pi/2c$, as shown in Figs.~1 and \ref{fig:BETS-sym}. At the same time, as shown in Supplemental Material~\ref{sec:g_perp.symmetry} and illustrated in the insets of Figs.~1 and \ref{fig:BETS-sym}, $g_\perp ({\bf k})$ is odd under reflection around the same line. As a result, for the $\delta$ pocket $\bar{g}_\perp$ in Eq.~(\ref{eq:g.average}) vanishes, as stated in the main text.

\subsection{Symmetry analysis of the N\'eel state of
$\text{Nd}_{2-x}\text{Ce}_x\text{CuO}_4$ in a transverse field}
\label{Sec:NCCO-symmetry}

In the antiferromagnetic state of $\text{Nd}_{2-x}\text{Ce}_x\text{CuO}_4$ (hereafter NCCO), the Cu$^{2+}$ spins point along the layers. At zero field, they form a so-called non-collinear structure: the staggered magnetization vectors of adjacent layers are normal to each other, pointing along the crystallographic directions $[100]$ and $[010]$, respectively (see Ref.\,\cite{armi10} for a review). However, an in-plane field above 5\,T transforms this spin structure into a collinear one, with the staggered magnetization in all the layers aligned transversely to the field. Therefore, in our experiment, with the field $B > 45$\,T rotated around the $[100]$ axis, the staggered magnetization is normal to the field at all tilt angles except for a narrow interval $0^{\circ} < |\theta| \lesssim 5^{\circ}$.

\begin{figure}[tb] 
\center
\includegraphics[width = .48 \textwidth]{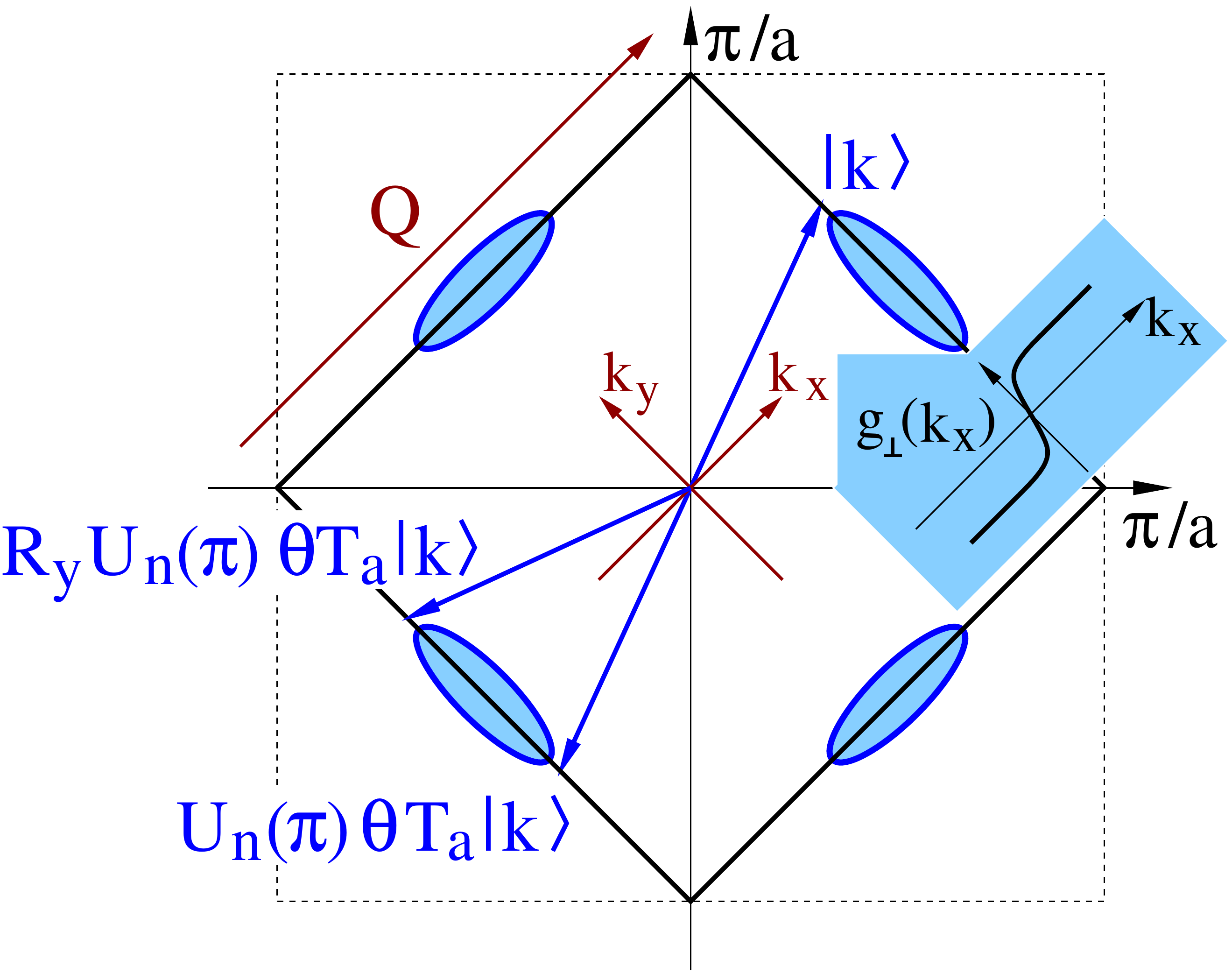}
\caption{\textbf{Schematic view of the BZ of NCCO.} The BZ in its paramagnetic state (dashed line) and in the N\'eel state with wave vector ${\bf Q} = (\pi/a , \pi/a)$ (solid line). The carrier pockets $\alpha$ are centered at the midpoints ${\bf k} = (\pm \pi/2a , \pm \pi / 2a) $ of the magnetic BZ boundary. The blue arrows show an exact Bloch eigenstate $| {\bf k} \rangle$ at a wave vector ${\bf k}$ on the magnetic BZ boundary -- and its symmetry partners ${\bf U}_{\bf n} (\pi) {\theta} {\bf T}_a | {\bf k} \rangle$ and
${\bf R}_y {\bf U}_{\bf n} (\pi) {\theta} {\bf T}_a | {\bf k} \rangle$. The orthogonality $\langle {\bf k} | {\bf R}_y {\bf U}_{\bf n} (\pi) {\theta} {\bf T}_a | {\bf k} \rangle = 0$ implies that $g_\perp ({\bf k})$ vanishes on the segment $k_x a = \pm \pi/\sqrt{2}$. The inset illustrates $g_\perp ({\bf k})$ being an odd function of $k_x a - \pi/\sqrt{2}$.}
\label{fig:NCCO-sym}
\end{figure}

Thus, we can restrict ourselves to the purely transverse-field geometry, with the field normal to the N\'eel axis, which makes the analysis similar to that for $\kappa$-BETS.
The only difference is that, given the tetragonal symmetry of NCCO, the triple pro\-duct $\hat{\bf U}_{\bf n} (\pi) \hat{\theta} \hat{\bf T}_a$ can now be combined with reflections $\hat{\bf R}_x$: $(k_x , k_y) \rightarrow (-k_x , k_y)$ and $\hat{\bf R}_y$: $(k_x , k_y) \rightarrow (k_x , -k_y)$. As a result, for any wave vector ${\bf k}$ at the magnetic Brillouin-zone boundary, one finds $\langle {\bf  k} | \hat{\bf R}_y \hat{\bf U}_{\bf n} (\pi) \hat{\theta} \hat{\bf T}_a | {\bf  k} \rangle = \langle {\bf  k} | \hat{\bf R}_x \hat{\bf U}_{\bf n} (\pi) \hat{\theta} \hat{\bf T}_a | {\bf  k} \rangle = 0$. This guarantees double degeneracy of Bloch eigenstates, hence the equality $g_\perp ({\bf k}) = 0$ at the entire boundary of the magnetic BZ, as shown in Fig.~\ref{fig:NCCO-sym}.

The charge-carrier pockets of our interest are believed to be centered at $(\pm \pi/2a , \pm \pi/2a)$, and are symmetric with respect to reflections ${\bf R}_x$ and ${\bf R}_y$ around the $k_x$ and $k_y$ axes, as shown in Figs.~4 and \ref{fig:NCCO-sym}. At the same time, as shown in Section \textbf{\ref{sec:g_perp.symmetry}} and illustrated in the inset of Figs.~4 and \ref{fig:NCCO-sym}, $g_\perp ({\bf k})$ is odd under the very same reflections. As a result, $\bar{g}_\perp$ in Eq.~(\ref{eq:g.average}) vanishes for these pockets, as stated in the main text.

\section{Estimating the product $k_F \xi$}
\label{sec:kxi}

\subsection{$\delta$ pocket in $\kappa$-BETS}
Looking only for a crude estimate, we assume a parabolic energy dispersion and treat the $\delta$ pocket as circular of radius $k_F$ and area $\mathcal{F}_\delta = 2\pi e F_{\delta}/\hbar$. Defining the antiferromagnetic coherence length as $\xi = \hbar v_F / \Delta_{\mathrm{AF}}$, we find:
\be
\label{eq:kFxi}
k_F \xi = \hbar k_F v_F / \Delta_{\mathrm{AF}} = 2 \varepsilon_F / \Delta_{\mathrm{AF}} .
\ee
The Fermi energy $\varepsilon_F$ in Eq.~(\ref{eq:kFxi}) can be expressed via the Shubnikov-de Haas (SdH) frequency  $F_\delta=61$\,T:
\begin{equation}
\label{eq:EF}
\varepsilon_F = \frac{\hbar^2 k_F^2}{2m} = \frac{\hbar^2 \mathcal{F}_\delta}{2\pi m} = \frac{\hbar e F_\delta}{m} \approx 6 \, \text{meV}.
\end{equation}
Assuming a BSC-like relation between the N\'{e}el temperature, $T_N \approx 2.5$\,K, and the antiferromagnetic gap $\Delta_{\mathrm{AF}}$ in the electron spectrum, we evaluate the latter as $\Delta_{\mathrm{AF}} \simeq 1.8 k_BT_N \approx 0.4$\,meV. A similar estimate is obtained from the critical field, $B_c \approx 5$\,T, required to suppress the  N\'{e}el state: $\Delta_{\mathrm{AF}} \sim \mu_B B_c \approx 0.3$\,meV.

Thus, we find $k_F \xi \simeq \frac{2 \varepsilon_F}{\Delta_{\mathrm{AF}}} \sim 30-40 \gg 1$, which means that $g_\perp ({\bf k})$ is nearly constant over most of the Fermi surface, except in a small vicinity of $k_c = \pi/2c$, where it  changes sign, cf. Figs.~\ref{fig:BETS-sym} and \ref{fig:g1D}.

\subsection{Small hole pocket of the reconstructed Fermi surface in NCCO}

In NCCO, the small Fermi-surface pocket $\alpha$, responsible for the observed oscillations, is far from being circular. Therefore, we can no longer estimate $k_F \xi$ the same way as we did for the $\delta$ pocket in $\kappa$-BETS. Instead, we evaluate the relevant Fermi wave vector and the antiferromagnetic coherence length separately.

The value of the Fermi wave vector in the direction normal to the magnetic BZ boundary can be found from ARPES maps of the Fermi surface~\cite{he19,armi02,mats07}:  $k_F = 0.4 \pm 0.1$\,nm$^{-1}$.

The coherence length can be estimated using the MB gap value, $\Delta_{\mathrm{AF}} \approx 16$\,meV, and parameters of the (approximately circular) large parent Fermi surface obtained from the analysis of MB quantum oscillations~\cite{helm15}. Using the corresponding SdH frequency $F = 11.25$\,kT and cyclotron mass $m_c = 3.0 m_e$, we estimate the Fermi velocity, $v_F \sim \hbar k_F /m_c \approx \sqrt{2\hbar eF}/m_c \approx 2.2 \times 10^5$\,m/s, which leads to the coherence length $\xi \sim \hbar v_F/ \Delta _{\mathrm{AF}} \approx 9$\,nm.

This yields the product $k_F \xi\sim 3-5$, which implies $g_\perp ({\bf k})$ being piecewise nearly constant over most of the Fermi surface, except in a small vicinity of the magnetic BZ boundary, where $g_\perp ({\bf k})$ changes sign, cf. Figs.~\ref{fig:NCCO-sym} and \ref{fig:g1D}.

\section{Symmetry properties of $g_\perp ({\bf k})$}
\label{sec:g_perp.symmetry}

In Section~\textbf{\ref{Sec:BETS-symmetry}}, we have shown that in $\kappa$-BETS the factor $g_\perp ({\bf k})$ vanishes at the entire $k_c = \pm \pi/2c$ segment of the magnetic BZ boundary. Here, we establish an important general symmetry property of $g_\perp ({\bf k})$. In the case of $\kappa$-BETS, this property implies that $g_\perp ({\bf k})$ is an odd function of $k_c - \pi/2c$. The $\delta$ pocket,  responsible for the observed SdH oscillations, is centered on this segment, at the corner of the magnetic BZ (see Fig.~1). As a result, for this pocket the ``effective $g$-factor'' $\bar{g}_\perp$ in Eq.~(\ref{eq:g.average}) vanishes by symmetry.
\begin{figure}[b] 
	\center
	\includegraphics[width = .35 \textwidth]{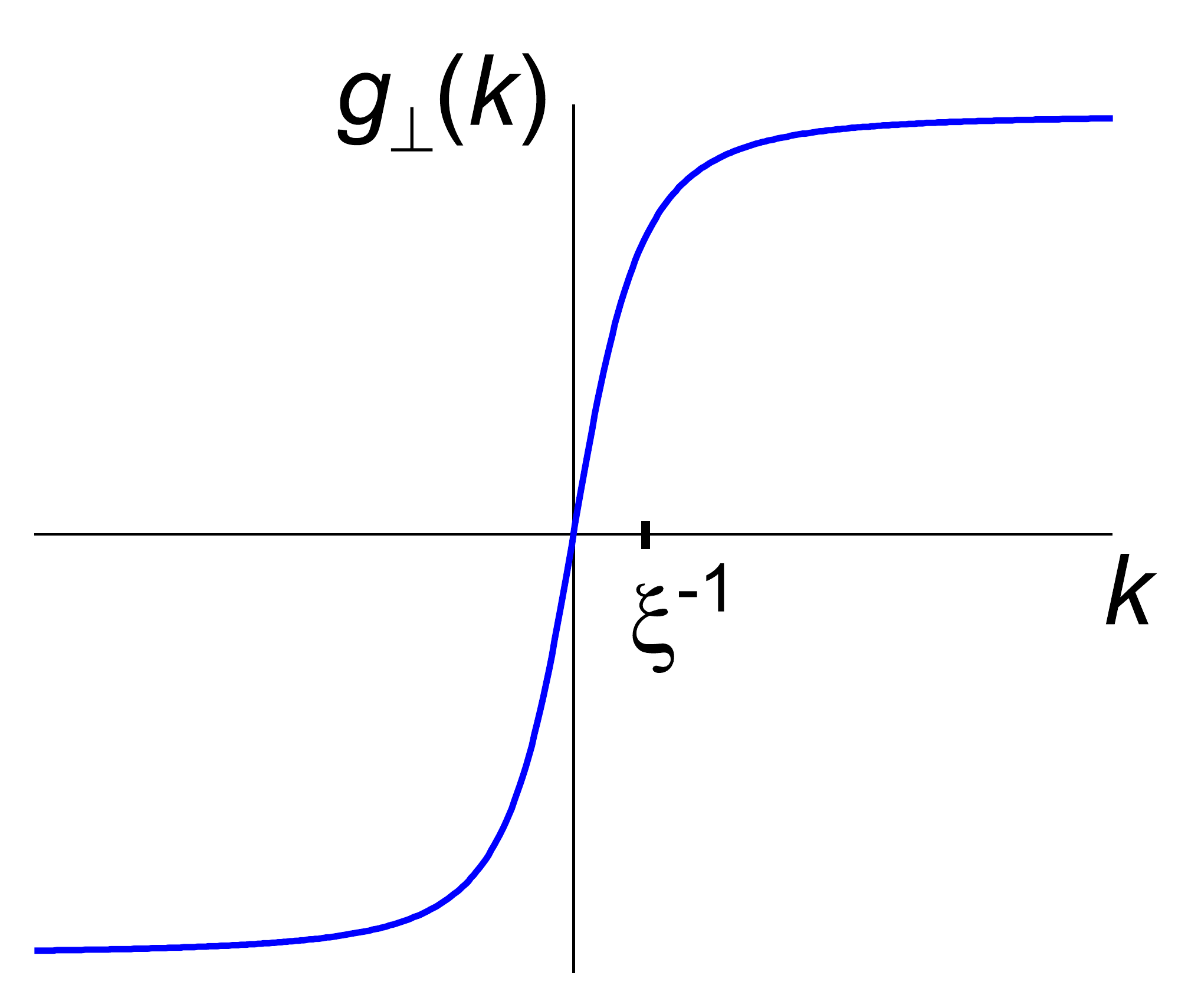}
	\caption{\textbf{The $g$-factor for magnetic field normal to the N\'eel axis.} Schematic plot of $g_\perp (k)$ as a function of the momentum component $k$, normal to the line $g_\perp ({\bf k}) = 0$. At small $k < 1/\xi$, the function $g_\perp (k)$ is linear: $g_\perp (k) \approx g \xi k$. Beyond $k \approx 1/\xi$, $g_\perp (k)$ is nearly constant: $g_\perp (k) \approx g$. Here, $\xi = \hbar v_F/\Delta_{\text{AF}}$ is the antiferromagnetic coherence length, and $\Delta_{\text{AF}}$ is the energy gap in the electron spectrum (\ref{eq:spectrum}) of the N\'eel state.}
	\label{fig:g1D}
\end{figure}

Without loss of generality, we consider the simplest case of double commensurability, relevant to both materials of our interest. In both of them, the underlying non-magnetic state is centrosymmetric, with the relevant electron band having the spectrum $\varepsilon({\bf k})$. Spontaneous N\'eel magnetization with wave vector ${\bf Q}$ interacts with the conduction-electron spin ${\bm \sigma}$ via the exchange term $({\bm \Delta}_{\text{AF}} \cdot {\bm \sigma})$, coupling the states at wave vectors ${\bf k}$ and ${\bf k + Q}$. In the N\'eel phase, subjected to magnetic field ${\bf B}$, the electron Hamiltonian takes the form~\cite{KulTug.1984}
\begin{equation}
\label{eq:Hp}
\mathcal{H}_{\bf k} =
\left[
\begin{array}{cc}
\varepsilon({\bf k}) - g \left( {\bf B} \cdot {\bm \sigma} \right) &
({\bm \Delta}_{\text{AF}} \cdot {\bm \sigma}) \\
( {\bf \Delta}_{\text{AF}} \cdot {\bm \sigma}) &
\varepsilon({\bf k + Q}) - g \left( {\bf B} \cdot {\bm \sigma}
\right)
\end{array}
\right] ,
\end{equation}
where the factor $\mu_B / 2$ has been absorbed into the de\-fi\-ni\-tion of ${\bf B}$, and ${\bm \Delta}_{\text{AF}} = J {\bf S}$ is the product of the antiferromagnetically ordered
moment ${\bf S}$ and its exchange coupling $J$ to the conduction electrons. In a purely transverse field ${\bf B}_\perp \perp {\bm \Delta}_{\text{AF}}$, the Hamiltonian
(\ref{eq:Hp}) can be easily diagonalized~\cite{RevazPRB2009-1,BraLuRa1989} to yield the spectrum
\be
\label{eq:spectrum}
\mathcal{E}({\bf k})
 =
 \varepsilon_+({\bf k}) \pm \sqrt{\Delta_{\text{AF}}^2 + \left[\varepsilon_-({\bf k})
  - g \left( {\bf B}_\perp  \cdot {\bm \sigma} \right) \right]^2} ,
 \ee
where $\varepsilon_\pm({\bf k}) = \frac{1}{2} \left[ \varepsilon({\bf k}) \pm \varepsilon({\bf k + Q}) \right]$. Equation~(\ref{eq:spectrum}) shows that $\Delta_{\text{AF}}$ is the energy gap in the electron spectrum of the N\'eel state. From Eq.~(\ref{eq:spectrum}), one easily finds the effective transverse $g$-factor $g_\perp({\bf k})$~\cite{RevazPRB2009-1,KabanovAlexandrov}
\be
\label{eq:model.g_perp}
g_\perp({\bf k}) = \frac{g \varepsilon_-({\bf k})}{\sqrt{\Delta_{\text{AF}}^2 + \varepsilon^2_-({\bf k})}} ,
\ee
plotted in Fig.~\ref{fig:g1D} as a function of momentum component $k$, normal to the line $g_\perp({\bf k}) = 0$.

The parent paramagnetic state is invariant under time reversal, thus $\varepsilon({\bf k}) = \varepsilon(-{\bf k})$. Also, in a doubly-commensurate antiferromagnet with N\'eel wave vector ${\bf Q}$, the wave vector $2{\bf Q}$ is a reciprocal lattice vector of the underlying non-magnetic state; thus, $\varepsilon({\bf k} + 2 {\bf Q}) = \varepsilon({\bf k})$. From these properties, it follows that $\mathcal{E}({\bf k}) = \mathcal{E}(- {\bf k} + {\bf Q})$ and $g_\perp({\bf k}) = - g_\perp(-{\bf k}+ {\bf Q})$~\cite{RevazPRB2009-1}. We will now show how this symmetry property leads to $\bar{g}_\perp = 0$. In NCCO as well as in the N\'eel state of $\kappa$-BETS, the relevant Fermi surface consists of two symmetric parts, which map onto each other under transformation ${\bf k} \rightarrow  - {\bf k} + {\bf Q}$. Contributions of these two parts to the integral in the right-hand side of Eq.~(\ref{eq:dF}) cancel each other exactly; hence, Eq.~(\ref{eq:g.average})
yields $\bar{g}_\perp = 0$. In other words, Eq.~(\ref{eq:dF}) yields $\delta\mathcal{F} = 0$, and thus, in Eqs.~(2) and~(3) one finds $R_s(\theta) = 1$: in a transverse field, the amplitude
of magnetic quantum oscillations has no spin zeros.

The arguments above rely on a quasi-\-clas\-si\-cal description. Note that the key conclusion, the absence of spin zeros in a transverse field, holds regardless of how the Fermi wave vector $k_F$ compares with the inverse antiferromagnetic coherence length $1/\xi \sim \Delta_{\text{AF}} / \hbar v_F$, where the behavior of $g_\perp ({\bf k})$ crosses over from linear  to constant as illustrated in Fig.~\ref{fig:g1D}.

In the limit of $k_F \xi \lesssim 1$, the problem can be analyzed by reducing the Hamiltonian to the leading terms of its momentum expansion around the band extremum. The conclusion remains intact: in a purely transverse field, the Zeeman term of Eq.~(1) does \textit{not} lift the spin degeneracy of Landau levels~\cite{RevazPRB2009-2}; hence,  the quantum-oscillation amplitude has no spin zeros. The present work extends the validity range of this result from a small Fermi-surface pocket to an arbitrarily large Fermi surface.

\section{Quantum oscillations in ${\bf{\mathrm CeIn}_3}$}
\label{Sec:CeIn3}

\begin{figure}[tb] 
\centerline{
   \mbox{\includegraphics[width=1.7in]{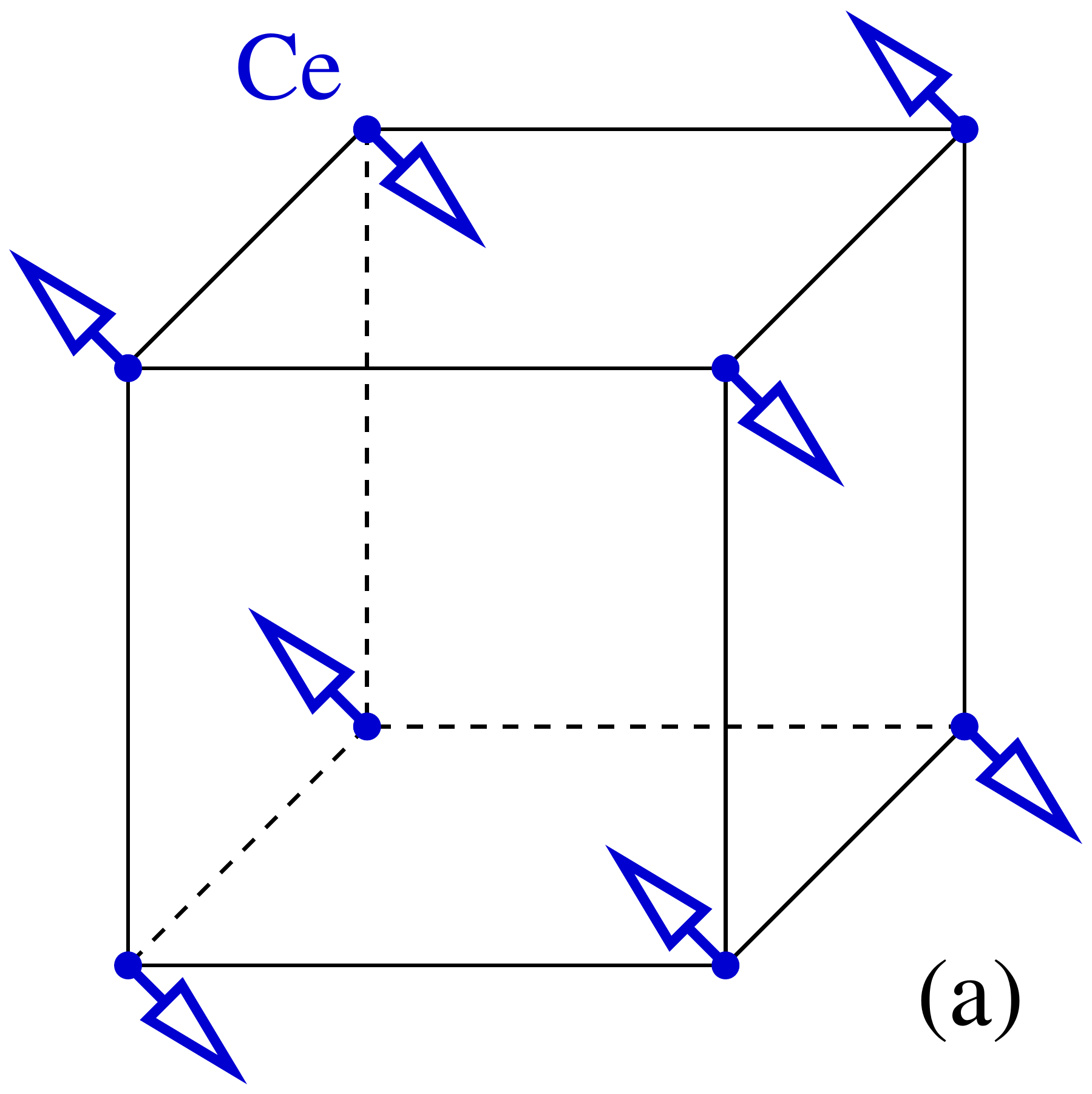}}
 \hspace{0cm}
   \mbox{\includegraphics[width=1.9in]{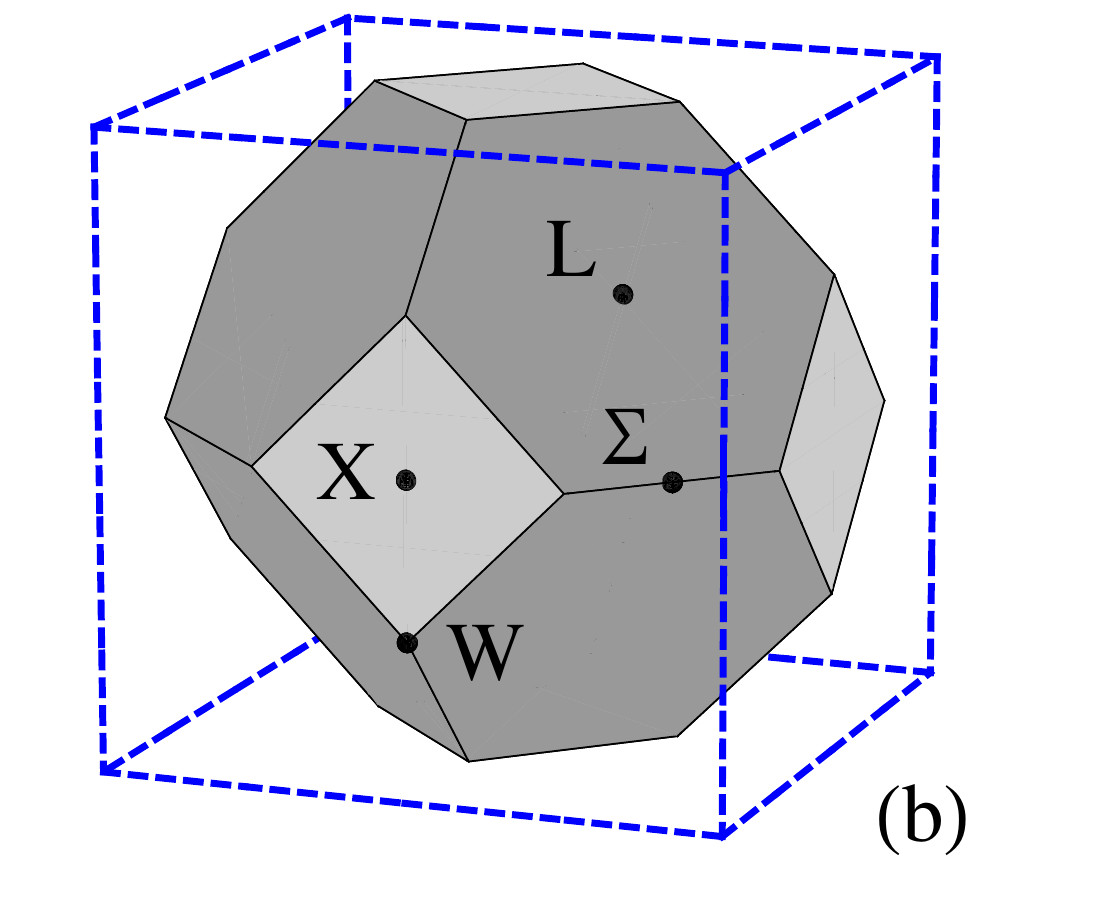}}
 \hspace{-0cm}
 }
 \caption{\textbf{Geometry of CeIn$_3$ in real and reciprocal space.} (a) Cubic unit cell of CeIn$_3$ in its N\'eel state, showing Ce atoms and their magnetic moments~\cite{Lawrence-80}. Indium atoms (not shown) are located at the face centers of the unit cell. (b) Cubic BZ of paramagnetic CeIn$_3$ shown by dashed lines and, inside, the magnetic BZ. Hexagonal faces (dark gray) form the reciprocal-space surface of the symmetry-protected degeneracy $g_\perp ({\bf k}) = 0$~[\onlinecite{RevazPRB2009-1}].
}
 \label{fig:CeIn3}
\end{figure}
\begin{figure}[!h] [H]
	\center
	\includegraphics[width = .3 \textwidth]{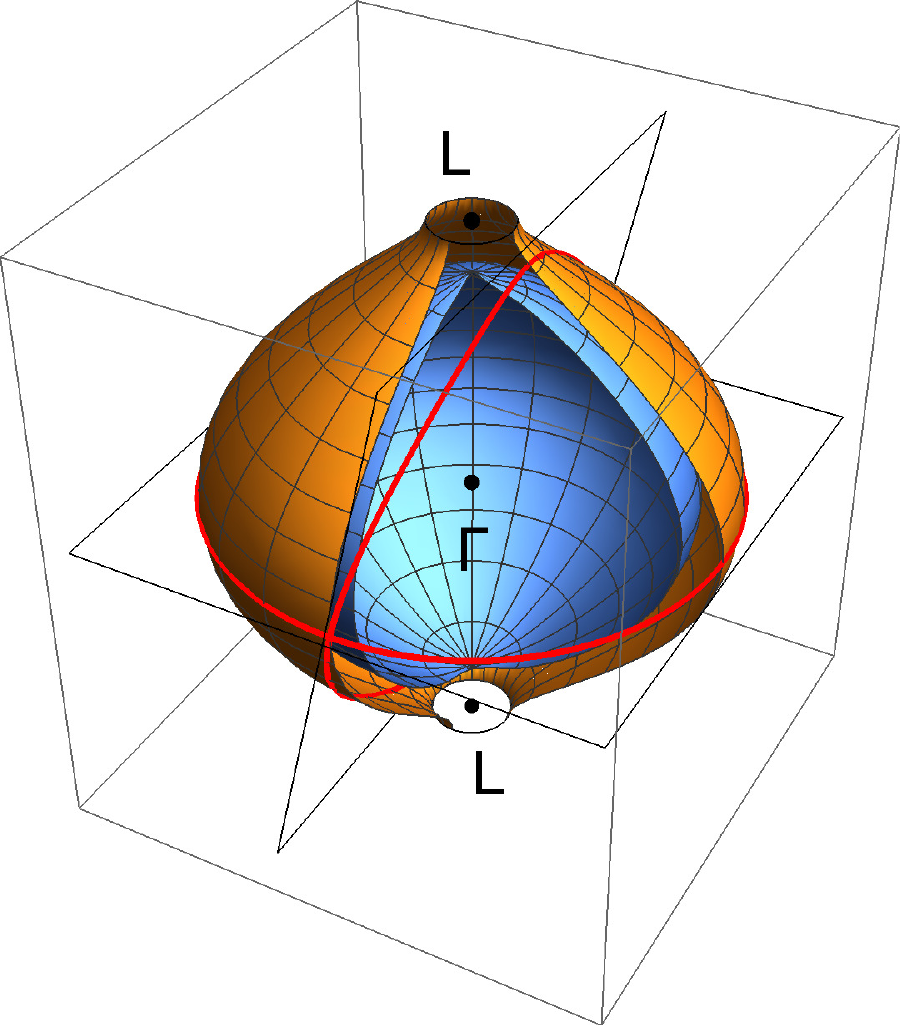}
	\caption{\textbf{Schematic view of the $d$ sheet of CeIn$_3$ in the N\'eel state.} We only show a single pair of necks for clarity. The outer sheet illustrates the Fermi-surface reconstruction with broadened necks, the inner sheet shows a reconstruction with the necks truncated. The red lines represent two cyclotron trajectories: a smaller cyclotron-mass trajectory in a magnetic field pointing along $\Gamma$L, and a larger cyclotron-mass trajectory, approaching the neck of the Fermi surface.}
	\label{fig:CeIn3FS}
\end{figure}
As we pointed out in the main text, for certain Fermi surfaces Zeeman spin-orbit coupling may not manifest itself in quantum oscillations: this is sensitive to where the Fermi surface is centered. An illustrative example is provided by CeIn$_3$, a heavy-fermion compound of simple-cubic Cu$_3$Au structure, with a moderately enhanced Sommerfeld coefficient, $\gamma= 130\,$mJ$/$K$^2\,$mol. Below $T_N \approx 10\,$K, it develops a type-II antiferromagnetic structure with wave vector ${\bf Q} = ( \frac{\pi}{a} , \frac{\pi}{a} , \frac{\pi}{a} )$ and an ordered moment of about (0.65 $\pm$ 0.1) $\mu_B$ per Ce atom~\cite{Lawrence-80}, shown in Fig.~\ref{fig:CeIn3}(a). The material remains a metallic down to the lowest temperatures. The BZs in the paramagnetic and in the N\'eel state are shown in Fig.~\ref{fig:CeIn3}(b). In a transverse magnetic field, anti-unitary symmetry protects double degeneracy (and thus the equality $g_\perp ({\bf k}) = 0$) on hexagonal faces of the magnetic BZ in Fig.~\ref{fig:CeIn3}(b)~\cite{RevazPRB2009-1}.

We will focus on the $d$ branch of the Fermi surface of CeIn$_3$, a nearly spherical sheet centered at the zone center $\Gamma$, with a radius close to $\frac{\sqrt{3}}{2} \frac{\pi}{a}$, where $a$ is the lattice constant. In the paramagnetic state, this sheet has necks protruding out near the $L$ points in Fig.~\ref{fig:CeIn3}(b), connecting it
to another sheet centered at the corner of the paramagnetic BZ~\cite{Biasini-03,Rusz-05}.

The $d$ sheet has been studied in detail by quantum oscillations, which revealed a large enhancement of the cyclotron mass upon tilting the field, from $m \approx 2 m_0$ for cyclotron trajectories passing far from the necks (e.g., for magnetic field ${\bf B} \| \langle 1 0 0 \rangle$) to $m > 12 m_0$ for trajectories approaching the necks (such as ${\bf B} \| \langle 1 1 0 \rangle$)~\cite{Settai-95,Ebihara-04}. This mass enhancement can be explained~\cite{Gorkov-Grigoriev-06} by the Fermi-surface geometry, illustrated in Fig.~\ref{fig:CeIn3FS}: approaching a neck, the cyclotron trajectory inevitably runs into a saddle point with its concomitant logarithmic enhancement of the cyclotron mass.

Upon transition to the N\'eel state, the Fermi surface undergoes a reconstruction by folding into the magnetic BZ. Depending on the neck size and on the value of the N\'eel gap in the electron spectrum, the reconstructed sheets may have their necks broadened or truncated altogether~\cite{Gorkov-Grigoriev-06}, as shown in Fig.~\ref{fig:CeIn3FS}.

The cyclotron-mass enhancement for those field orientations, for which the quasiclassical trajectories approach a neck~\cite{Settai-95,Ebihara-04} is consistent with the Fermi sheet having necks in the N\'eel state. So is the observation of spin zeros~\cite{Settai-95}: the enhancement of $m/m_0$ alone produces spin zeros as the cyclotron trajectory approaches a neck with tilting the field. Crucially for the Zeeman effect that we are interested in, the average of $g_\perp ({\bf k})$ over a cyclotron trajectory on the $d$ sheet does \textit{not} vanish -- simply because most (if not all) of the cyclotron trajectory in question lies \textit{within} the first magnetic BZ, and thus $g_\perp ({\bf k})$ keeps its sign over most (or all) of the trajectory.
Indeed, being centered at the $\Gamma$ point, the $d$ sheet cannot possibly be symmetric with respect to the surface $g_\perp ({\bf k}) = 0$ in Fig.~\ref{fig:CeIn3}(b), and thus our symmetry argument for the vanishing $\bar{g}_\perp$ inevitably breaks down. The observation of spin zeros~\cite{Settai-95} in CeIn$_3$ is thus perfectly consistent with the physics of the Zeeman spin-orbit coupling described in the main text.

\section{$\text{SdH}$ oscillations in the high-field paramagnetic state of $\kappa$-\text{BETS}} \label{sec:SdH-BETS}

The SdH oscillations in the high-field, paramagnetic (PM) state of $\kappa$-BETS have been described in detail in a number of publications~\cite{bali00,uji01c,kono05,kono06,kart16}, revealing a Fermi surface largely consistent with that obtained from band-structure calculations~\cite{fuji01}. Here, we give a brief overview, referring to our own data, which show perfect agreement with the previous reports. The measurements  were performed on the same crystal as discussed in the main text.

Figure\,1(b) in the main text shows an example of the low-temperature interlayer resistance measured in a magnetic field up to 14\,T, directed nearly perpendicularly to the conducting layers. The kink around $B_c \approx 5.2$\,T reflects the transition from the low-field AF to the high-field PM state. The slow SdH oscillations associated with the small pocket $\delta$ (shown in Fig. 1c) of the reconstructed FS in the AF state collapse at $B_c$. In the PM state, the oscillation spectrum is composed of two fundamental frequencies $F_{\alpha} \approx 840$\,T and $F_{\beta} \approx 4200$\,T and their combinations, as shown in the inset in Fig.1(b). The dominant contribution comes from the $\alpha$ oscillations associated with the closed portion of the Fermi surface centered on the BZ boundary shown in Fig.\,1(a). The $\beta$ oscillations, which are considerably weaker than the $\alpha$ oscillations, originate from the large magnetic-breakdown (MB) orbit enveloping the whole Fermi surface [dashed orange line in Fig.\,1(b)]. Additionally, there are sizable contributions of the second harmonic of the $\alpha$ oscillations and combination frequencies, $\beta - p\alpha$ with $p = 1, 2$, in the SdH spectrum, which are often observed in clean, highly two-dimensional organic metals in the intermediate MB regime~\cite{kart04}. 
	The oscillation amplitude shows spin-zeros not only in the angular dependence but also with changing the field strength at a fixed orientation. This is due to the presence of an exchange field $B_J$ imposed on the conduction electrons by saturated paramagnetic Fe$^{3+}$ ions, although some details of its behavior are still to be understood~\cite{shoe84,cepa02,kart16}. The analysis of the spin-zero positions in the PM state yields the $g$-factor very close to that of free electrons, $g = 2.0 \pm 0.2$ and the exchange field $B_J \approx -13$\,T pointing against the external magnetic field \cite{kart16}.

Note that both $\alpha$ and $\beta$ oscillations are rapidly suppressed with lowering the field and cannot be resolved near the transition field $B_c$. This rapid suppression is due to the relatively high cyclotron masses: compared to the $\delta$-oscillation mass, the $\alpha$ and $\beta$ masses are some $5$ and $7$ times higher, respectively. The exponential dependence of the SdH amplitude on the cyclotron mass, see Eq.~(3) of the main text, renders the amplitude factor for the $\alpha$ oscillations approximately 2 orders of magnitude smaller than that for the slow $\delta$ oscillations at fields near $B_c$. In addition to the higher cyclotron mass, the MB gap $\Delta_0$ contributes to the damping of the $\beta$ oscillations via the MB damping factor. This is why the $\beta$ oscillations are only seen at highest fields, $\geq 10$\,T, as weak distortions of the $\alpha$-oscillation wave form and as a small peak in the FFT spectrum.

\section{Details of the $\text{SdH}$ fit for $\kappa$-\text{BETS} in the AF state} \label{sec:noeffect}

In the main text, we noted a large uncertainty of $B_0$ and $B_{AF}$. Here, we show that the quality of our SdH amplitude fits is insensitive to the exact values of $B_0$ and $B_{AF}$. Equation~(3) for the amplitude $A_{\delta}$ contains the MB factor
\begin{equation}
R_{\mathrm{MB}}^{[\delta]} = \left[1-\exp\left(-\frac{B_0}{B\cos\theta}\right)\right] \left[1-\exp\left(-\frac{B_{\mathrm{AF}}}{B\cos\theta}\right)\right],
\label{BETS_MB}
\end{equation}
which must be taken into account when analyzing the angular dependence $A_{\delta}(\theta)$.

A rough estimate of $B_0$ can be obtained from the ratio between the $\alpha$- and $\beta$-oscillation amplitudes at a certain field and temperature, using the LK formula [Eq.~(3) of the main text] with the MB factors $R_{\mathrm{MB}}^{[\alpha]} = \left[1-\exp\left(-B_0/B\right)\right]$ and $R_{\mathrm{MB}}^{[\beta]} = \exp\left(-2B_0/B\right)$ for the $\alpha$ and $\beta$ oscillations, respectively. From the data in Fig.\,1(b) we estimate the ratio between the FFT amplitudes of the $\alpha$ and $\beta$ oscillations in the field interval 10 to 14\,T, at $T = 0.5$\,K, as $A_{\beta}/A_{\alpha} \approx 0.03$. To complete the estimation, we also need to know the cyclotron masses, which have been determined in earlier experiments: $m_{\alpha} = 5.2 m_0$ and $m_{\beta} = 7.9 m_0$~\cite{uji01c}, and the Dingle temperature $T_\mathrm{D}$. The latter arises from scattering on crystal imperfections~\cite{shoe84} and, therefore, varies from sample to sample. As shown below, for the present crystal a reasonable estimate of the Dingle temperature in the AF state, $T_\mathrm{D} \simeq 0.7$\,K, can be obtained from the angular dependence of the $\delta$-oscillation amplitude. Assuming that $T_\mathrm{D}$ is momentum-independent and remains the same in the PM state, we substitute this value in the LK formula, arriving at the MB field value $B_0 \approx 12$\,T. This is three times higher than the upper end of the field interval used for the SdH oscillation analysis in the AF state. Therefore, the first factor in the right-hand side of Eq.~(\ref{BETS_MB}) is close to unity and does not contribute significantly to the angular dependence $A_{\delta}(\theta)$~\cite{comm_TD}. To confirm this, we have checked how our fits are affected by varying $B_0$ in the range between 5\, T and 50\,T, as will be presented below.

The MB field $B_{\mathrm{AF}}$ is due to magnetic ordering and can be estimated from the gap $\Delta_{\mathrm{AF}}$ with the help of the Blount criterion ~\cite{shoe84,blou62},
\begin{equation}
B_{\mathrm{AF}} \sim \frac{m_c}{\hbar e} \cdot \Delta_{\mathrm{AF}}^2/\varepsilon_F \simeq 0.15\,\text{T}\,.
\end{equation}
Here, we estimated the Fermi energy $\varepsilon_F$ from SdH oscillations in the paramagnetic state: $\varepsilon_F \sim \hbar^2 k_F^2/2m \sim \hbar e F_{\beta}/m_{c,\beta}$, with the SdH frequency $F_{\beta} = 4280$\, T and corresponding cyclotron mass $m_{c,\beta} = 7.9 m_0$~\cite{uji01c}.

Of course, these are only rough estimates. Moreover, the observation of the $\delta$ oscillations in fields up to $B_c\simeq 5$\,T implies that the relevant MB field must be in the range of a few tesla, to provide a non-vanishing second factor in the right-hand side of Eq.~(\ref{BETS_MB}). 
On the other hand, the MB field cannot be much higher than the fields we applied ($B \lesssim B_c$).
We have tentatively set the upper estimate for $B_{\mathrm{AF}}$ to about 5\,T and checked how our fits are affected by varying $B_{\mathrm{AF}}$ from 0.15\, T to 5\,T.
\begin{figure}[tb]
	\centering
	\includegraphics[width=0.45\textwidth]{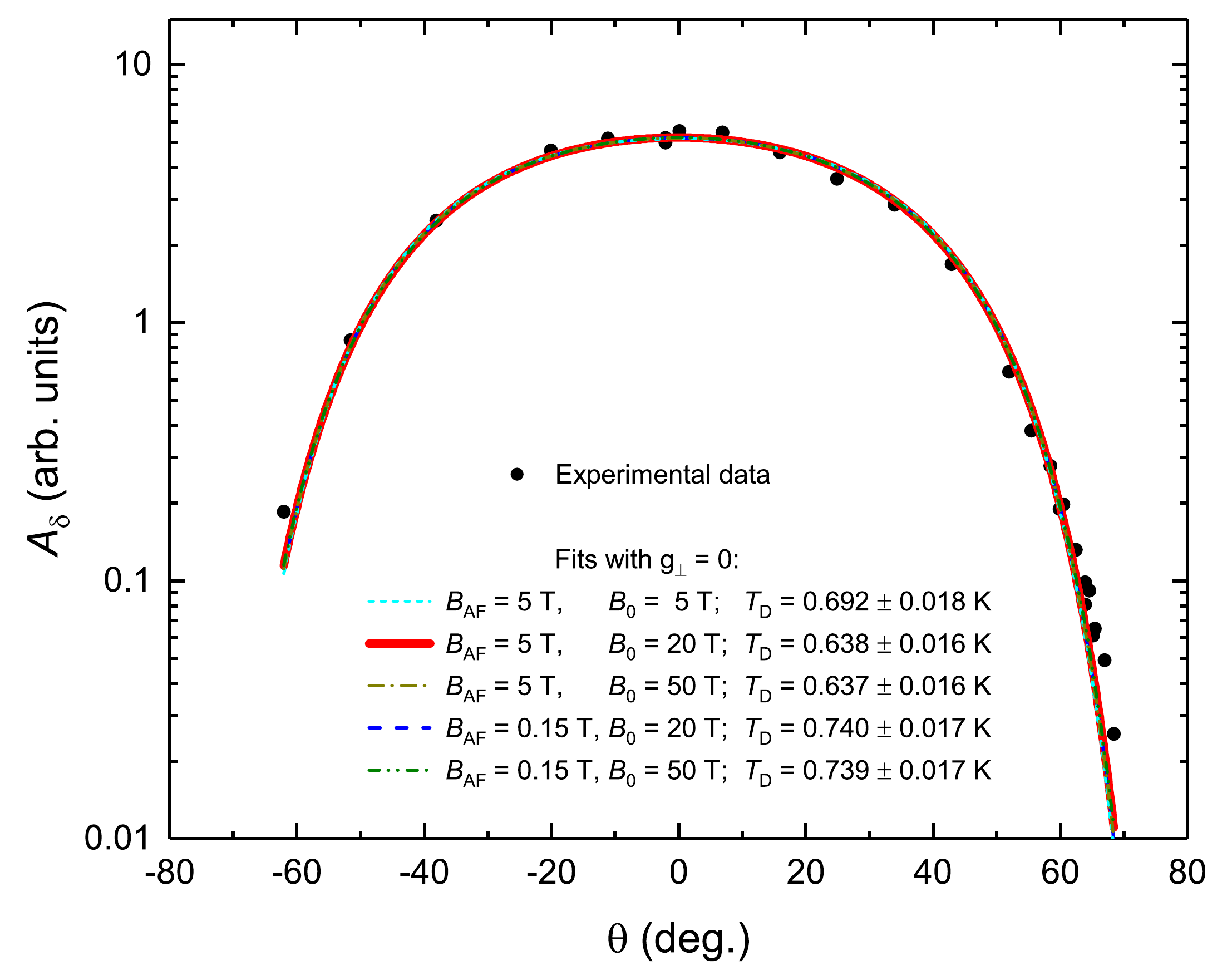}
	\caption{{\bf Angle-dependent $\delta$ oscillation amplitude in $\kappa$-BETS compared to theoretical fits.} Black dots are the experimental data and lines are fits using Eq. (3) of the main text with different fixed values of the MB fields $B_0$ and $B_{\mathrm{AF}}$. The angle-independent Dingle temperature $T_\mathrm{D}$ and amplitude prefactor $A_0$ are the fitting parameters and $g_{\perp}$ is set to zero. The plot demonstrates insensitivity of the angular dependence on the concrete choice of $B_0$ and $B_{\mathrm{AF}}$.
	}
	\label{A_theta_calc}
\end{figure}

The results are summarized in Fig.~\ref{A_theta_calc}, which presents several  fits to the experimental data for $\kappa$-BETS (the same as in Fig.~3 of the main text), with values of $B_0$ and $B_{\mathrm{AF}}$ varying in a broad range: $5\,\mathrm{T} \leq B_0 \leq 50\,\mathrm{T}$ and $0.15\,\mathrm{T} \leq B_{\mathrm{AF}} \leq 5\,\mathrm{T}$. All the fits assume $g_{\perp} =0$; as shown in the main text, a finite value for $g_{\perp}$ would simply lead to sharp spin zeros, insensitive to the monotonic $\theta$ dependence of $R_\mathrm{MB}$. One can clearly see that the fits are nearly indistinguishable and virtually insensitive to variation of $B_0$ within the given range, whereas the variation of $B_{\mathrm{AF}}$ barely results in a $15\%$ change of the Dingle temperature: $T_\mathrm{D} = (0.69\pm 0.05)$\,K. The parameter $A_0$ in Eq.~(3) changes roughly in inverse proportion to $B_{\mathrm{AF}}$. However, $A_0$ is largely an empirical parameter, irrelevant to our study. Thus, we conclude that the mentioned uncertainty of the MB fields has no effect on the quality of our fits, as stated in the main text.
\\
\section{Amplitude analysis for NCCO}
\label{NCCO:Amplitude_analysis}
\begin{figure*}[tb]
	\centering
	\includegraphics[width=1\textwidth]{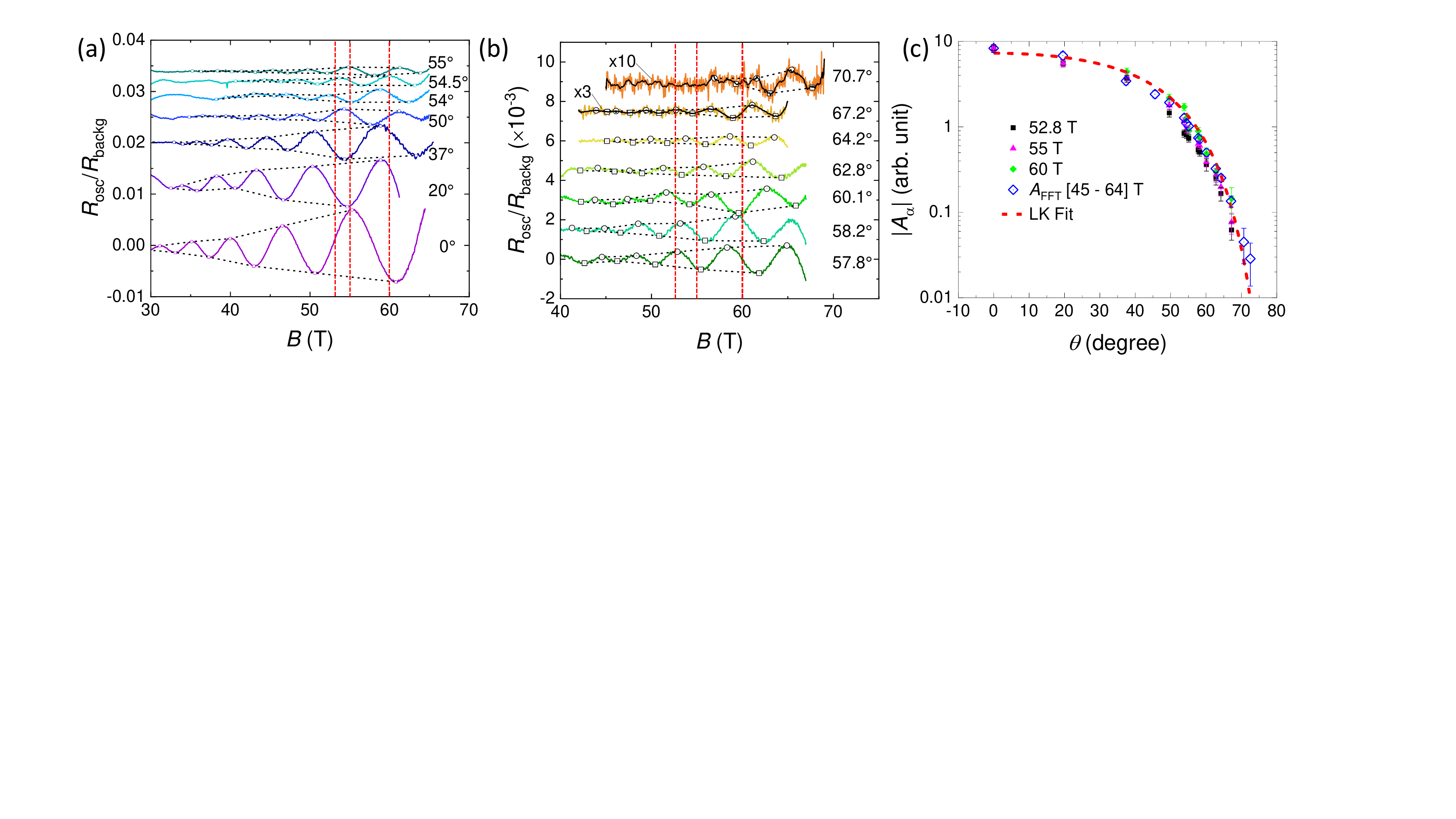}
	\caption{{\bf Amplitude analysis of the angle-dependent $\alpha$ oscillations in NCCO.} (a) and (b) oscillating resistance component, normalized to the non-oscillating $B$-dependent background, plotted as a function of magnetic field. The curves corresponding to different tilt angles $\theta$ are vertically shifted for clarity. For $\theta = 67.2^{\circ}$ and $70.7^{\circ}$ the ratio $R_{\text{osc}}/R_{\text{backg}}$ is multiplied by a factor of 3 and 10, respectively. The vertical dashed lines indicate the field values at which we determine $A(\theta)$ shown in (c). (c) Normalized peak-to-peak oscillation amplitude determined for $B=52.8$\,T, 55\,T, and 60\,T in comparison to the amplitude determined by FFT for $B=[45-64]\,$T. Red dashed line is the fit with $g=0$, see Fig.\,5 in the main text.
	}
	\label{NCCO_Amplitude_by_hand}
\end{figure*}

In our experiment on NCCO we were restricted to a relatively narrow field interval, between 45 and 64\,T where the oscillations were observable over the whole angular range. Due to the low frequency, this interval contains only a few oscillation periods, see Fig.~\ref{NCCO_Amplitude_by_hand} and Fig. 4 in the main text. Under these conditions the FFT spectrum is sensitive to the choice of the field window and details of the background subtraction. The choice of the upper bound of the field window is most important here: this affects the largest-amplitude oscillation and thus gives the dominant contribution to the overall error. In the course of our analysis we always ensured that the order of the polynomial fit ($\leq 4$) to the non-oscillating $B$-dependent background did not affect the height of the FFT peak corresponding to the SdH oscillations, within our resolution. The main source of error is the variation of the oscillation phase at the (fixed) upper end of the field window caused by the changing frequency $F(\theta)$: this disturbs the shape and amplitude of the largest oscillation. In order to reduce this parasitic effect, we kept the FFT window more narrow than that of the background fit; the upper $B$ border of the FFT window was at least 1\,T lower than that of the background fit. In our analysis we found that the associated error bar for the height of the FFT peak did not exceed $\pm 10\,\%$. While this uncertainty may be important, for example, for an exact evaluation of the effective cyclotron mass, in our case, when the oscillation amplitude is expected to change an order of magnitude near a spin-zero, it is not significant. Only at highest angles, the error becomes comparable to the signal, which is related to the low signal-to-noise ratio at these conditions.

To crosscheck our FFT analysis we present an alternative amplitude analysis in Fig.~\ref{NCCO_Amplitude_by_hand}. For each $R_{\mathrm{osc}}/R_{\mathrm{backg}}$ curve, we determine the peak-to-peak amplitude from the linearly interpolated envelopes (dashed black lines in Fig.~\ref{NCCO_Amplitude_by_hand}a,b) at a fixed field value.  We chose the middle of the FFT window in $1/B$ scale, i.e., $B=[(1/45+1/64)/2]^{-1}$\,T$=52.8\,$T, as well as the values 55\,T and 60\,T, with oscillations discernible up to $67\,$°. Fig.~\ref{NCCO_Amplitude_by_hand}c presents these points plotted versus the tilt angle $\theta$ and compares them to the FFT values and LK fit from Fig.~3 shown in the main text. The extracted data from both approaches match very well. 
Clearly, our main conclusion holds, that is, the overall angular dependence exhibits no indication of a spin-zero effect for angles of up to at least 70\,°.

\section{Estimating the disparity of spin-up and spin-down tunneling amplitudes in NCCO}
\label{Sec:possibilities}

\begin{details}
When looking for alternative explanations to our experimental findings, let us recall that,  generally, the effective $g$-factor may depend on the field orientation. This dependence may happen to compensate that of the quasi-2D cyclotron mass, $m/\cos\theta$, in the expression~(2) for the spin-reduction factor $R_s$, and render the latter nearly isotropic, with no spin zeros. Obviously, such a compensation requires a strong Ising anisotropy [$g(\theta=0^{\circ}) \gg g(\theta=90^{\circ})$] -- as found, for instance, in the heavy-fermion compound URu$_2$Si$_2$, with the values $g_c = 2.65 \pm 0.05$ and $g_{ab} = 0.0 \pm 0.1$ for the field along and normal to the $c$ axis, respectively~\cite{Alta12,Bast-19}. However, this scenario is irrelevant to both materials of our interest: In $\kappa$-BETS, a nearly isotropic $g$-factor, within a $20\%$ deviation from the free-electron value 2.0, was revealed by a study of spin zeros in the paramagnetic state~\cite{kart16}. In NCCO, the conduction electron $g$-factor may acquire anisotropy via an exchange coupling to Nd$^{3+}$ local moments. However, the low-temperature magnetic susceptibility of Nd$^{3+}$ in the basal plane is some 5 times larger than along the $c$ axis~\cite{Hund89,Dali93}. Therefore, the coupling to Nd$^{3+}$ may only \underline{increase} $g_{ab}$ relative to $g_c$, and thereby only \underline{enhance} the angular dependence of $R_s$ rather than cancel it  out. Thus, we are lead to rule out a $g$-factor anisotropy of crystal-field origin as a possible reason behind the absence of spin zeros in our experiments. 

As follows from Eq.~(2), another possible reason for the absence of spin zeros is a strong reduction of the ratio $gm/2m_{\mathrm{e}}$. However, while \textit{some} renormalization of this ratio in me\-tals is commonplace, its dramatic suppression (let alone nullification) is, in fact, exceptional. Firstly, a vanishing mass would contradict $m/m_{\mathrm{e}} \approx 1$,  experimentally found in both materials at hand. On the other hand, a Landau Fermi-liquid renormalization~\cite{shoe84} $g \rightarrow g / (1 + G_0)$ would require a colossal Fermi-liquid parameter $G_0 \geq 10$, for which there is no evidence in NCCO, let alone $\kappa$-BETS with its already mentioned $g \approx 2$ in the paramagnetic state~\cite{kart16}. 

A sufficient difference of the quantum-oscillation amplitudes and/or cyclotron masses for spin-up and spin-down Fermi surfaces might also lead to the absence of spin zeros. Some heavy-fermion compounds show strong spin polarization in magnetic field, concomitant with a substantial field-induced difference of the cyclotron masses of the two spin-split subbands~\cite{Harr-98,McCo-05}. 
As a result, for quantum oscillations in such materials, one spin amplitude considerably exceeds the other, and no spin zeros are expected. Note that this physics requires the presence of a very narrow conduction band, in addition to a broad one. In heavy-fermion compounds, such a band arises from the $f$ electrons, but is absent in both materials of our interest. 

Another extreme example is given by the single fully polarized band in a ferromagnetic metal, where only one spin orientation is present, and spin zeros are obviously absent. Yet, no sign of ferro- or metamagnetism has been seen in either NCCO or $\kappa$-BETS. Moreover, in $\kappa$-BETS,the spin-zero effect has been observed in the paramagnetic state~\cite{kart16}, indicating that the quantum-oscillation amplitudes of the two spin-split subbands are comparable.
\end{details}

The Nd$^{3+}$ spins in the insulating layers of NCCO are polarized by strong magnetic field. Consequently one may wonder whether this spin polarization could render interlayer tunneling amplitudes for spin-up and spin-down different enough to lose spin zeros in the $c$-axis magnetoresistance oscillations. In the following we will show that this possibility can be ruled out. The ratio of interlayer the conductivities $\sigma _\uparrow$ and $\sigma _\downarrow$ can be crudely estimated
via the tunneling amplitudes $w_\uparrow$ and $w_\downarrow$ as
\begin{widetext}
\[
\frac{\sigma _{\uparrow }}{\sigma _{\downarrow }} 
\approx \frac{w_{\uparrow }^{2}}{ w_{\downarrow }^{2}}=
\exp \left( -\frac{2}{\hbar} \int dz\left[ \sqrt{2m\left( U\left(
z\right) -E_{F}-E_{Z}/2\right) }-\sqrt{2m\left( U\left( z\right)
-E_{F}+E_{Z}/2\right) }\right] \right) ,
\] 
\end{widetext}
where $U \left( z \right)$ is the tunneling potential and $E_{Z}$ the Zeeman splitting, responsible for the difference between $\sigma _\uparrow$ and $\sigma _\downarrow$. For a rough estimate, it suffices to replace $U \left( z \right)$ by a rectangular potential barrier of spatial width $d$, the unperturbed tunneling amplitude being $w=\exp \left( -d \sqrt{2m\left( U\left( z\right)
-E_{F}\right) }/\hbar \right)$. The expression in the exponent above can then be expanded in small $E_{Z}$ to yield
\begin{widetext}
\[
\frac{\sigma _{\uparrow }}{\sigma _{\downarrow }} \approx
\exp \left(
\frac{E_{Z}}{U-E_{F}}
\frac{\sqrt{2m\left( U-E_{F}\right) }d}{\hbar }
\right)
 = \exp \left[ - \frac{E_{Z}}{U-E_{F}} \ln w \right].
\]
\end{widetext}
To put in the numbers, notice that the deviation of the experimental data in Fig.~5 of the main text from the theoretical fit with $R_s = 1$ does not exceed 20\%. Assuming this deviation to be entirely due to an interference of unequal spin-up and spin-down oscillation amplitudes would imply $\sigma _\uparrow/\sigma _\downarrow \sim 20$. Let us estimate the $E_Z$ required for such a behavior. Recall that $w$ is essentially the ratio of a typical interlayer hopping amplitude $t_z \sim 10^{-2}$ eV to the in-plane Fermi scale $U-E_{F} \sim 1$ eV, and thus $\ln w \approx - 4.6$. The ratio $\sigma _\uparrow/\sigma _\downarrow \sim 20$ would then mean $E_Z \gtrsim 0.6$ eV: an unrealistic bound, which allows us to rule out this scenario.